# Direct observation of coupled geochemical and geomechanical impacts on chalk microstructural evolution under elevated $CO_2$ pressure. Part I.

Y. Yang,[1]* S. S. Hakim,[1] S. Bruns,[1] M. Rogowska,[1] S. Boehnert,[1] J.U. Hammel,[2] S. L. S. Stipp[1] and H. O. Sørensen[1]

[1] Nano-Science Center, Department of Chemistry, University of Copenhagen, Universitetsparken 5, DK-2100 Copenhagen, Denmark

[2] Helmholtz-Zentrum Geesthacht, Max-Planck-Straße 1, 51502 Geesthacht, Germany

Current address: Universität Bremen, Bibliothekstraße 1, 28359 Bremen, Germany

Corresponding author: yiyang@nano.ku.dk

**Abstract**

The dissolution of porous media in a geologic formation induced by the injection of massive amounts of $CO_2$ can undermine the mechanical stability of the formation structure before carbon mineralization takes place. The geomechanical impact of geologic carbon storage is therefore closely related to the structural sustainability of the chosen reservoir as well as the probability of buoyance driven $CO_2$ leakage through caprocks. Here we show, with a combination of *ex situ* nanotomography and *in situ* microtomography, that the presence of dissolved $CO_2$ in water produces a homogeneous dissolution pattern in natural chalk microstructure. This pattern stems from a greater apparent solubility of chalk and therefore a greater reactive subvolume in a sample. When a porous medium dissolves homogeneously in an imposed flow field, three geomechanical effects were observed: material compaction, fracturing and grain relocation. These phenomena demonstrated distinct feedbacks to the migration of the dissolution front and severely complicated the infiltration instability problem. We conclude that the presence of dissolved $CO_2$ makes the dissolution front less susceptible to spatial and temporal perturbations in the strongly coupled geochemical and geomechanical processes.

**Graphical abstract**

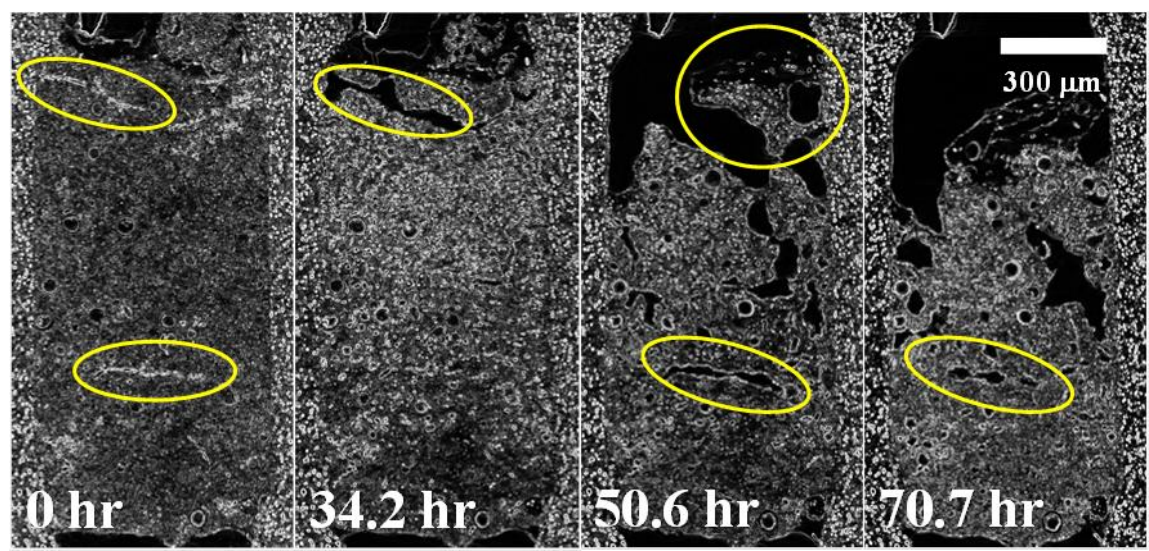

1. **Introduction**

The ratification of the Paris Agreement by the EU in November 2016 has once more pushed the global agenda for the implementation of geologic carbon storage (GCS) (Falkner, 2016). Concerns remain, however, regarding the societal and environmental consequences of GCS (DePaolo and Cole, 2013) – Is it safe? How much $CO_2$ can we bury? Both questions are closely related to the structural changes undergone by geologic formations after introducing concentrated $CO_2$. The opening or closing of permeable flowpaths determines whether buoyancy will drive the escape of sequestered $CO_2$ (Steefel et al., 2013). The microstructural evolution of porous formation rocks controls the outward migration of $CO_2$ from the injection well and therefore the efficiencies of solubility, residual and mineral trapping (Benson and Cole, 2008).

$CO_2$ dissolves in formation water and disrupts preexisting chemical equilibriums between water and rocks. This disequilibrium is the driving force for geochemical reactions which modify rock properties, such as mineral dissolution and precipitation. This modification, in turn, provides feedback on the migration of fluid that carries the reactants and products for these reactions (Yang et al., 2016a). This coupled evolution of flow field and microstructure through chemistry is far from being understood because of the difficulties in direct experimental observation (Noiriel, 2015) and in numerically handling mathematical models based on free boundary problems of partial differential equations (Chadam et al., 1986; Ortoleva et al., 1987; Szymczak and Ladd, 2012; Yang et al., 2016b). What further complicates the problem is the introduction of geomechanical effects (Jamtveit and Hammer, 2012; Keszthelyi et al., 2016; Røyne and Jamtveit, 2015). If a porous structure weakened by water-rock interactions cannot sustain the stress exerted by the formation and/or the flow field, it collapses and forms a new structure to reestablish mechanical stability. This constant renewal of microstructure again shapes the flow field and in long term determines the fate and transport of sequestered $CO_2$.

In this study we focus on the interplay between dissolved $CO_2$ and chalk, with an emphasis on the physical changes of microstructure in an imposed flow field. Chalk is predominantly composed of biogenic calcium carbonate produced by species of marine algae, either preserved as coccospheres with interlocked micrometre-scale platelets or reprecipitated as rhombohedral crystalline grains (Hassenkam et al., 2009). Chalk formations can be found in many localities, e.g., in the Gulf of Mexico or in the North Sea Basin, and serve as groundwater aquifers as well as oil and gas reservoirs (Hardman, 1982). In Nordic countries such as Denmark, chalk reservoirs are obvious candidates for GCS because of their availability and the associated possibilities of $CO_2$-enhanced oil recovery (D'Heur, 1984). Despite the large number of GCS-related publications on water-rock interactions over the last decade, studies on chalk remain very scarce, especially when it concerns the micro- and sub-micrometre origins of morphological evolution. Most insights on the potential $CO_2$ effects on chalk can only be derived from studies of carbonates from different geologic settings (Cao et al., 2016; Hao et al., 2013; Luquot et al., 2014; Noiriel et al., 2004; Noiriel et al., 2009; Noiriel et al., 2007; Polak et al., 2004; Wang and Tokunaga, 2015). The low content of silicates in chalk means the lack of cations for permanent mineralization of $CO_2$. As a result, GCS in chalk reservoirs relies primarily on three trapping mechanisms: stratigraphic trapping, solubility trapping and residual trapping (Benson and Cole, 2008; DePaolo and Cole, 2013). The Norwegian Sleipner site in North Sea has demonstrated the effectiveness of stratigraphic sealing in preventing the buoyancy-driven escape of $CO_2$ plume since the beginning of injection in 1996 (Zweigel et al., 2004). The solubility and residual trapping take place as the injected fluids, either supercritical $CO_2$ or water pre-equilibrated with concentrated $CO_2$, migrate outwards from the injection wells into the reservoir. This migration is determined by the generation and development of flow pathways in porous formation rocks (Zhao et al., 2015a, b). Accompanying this migration is a "wave" of cations – an increase in cation concentration (e.g., $[Ca^{2+}]$) that moves with the fluid flow. This wave is produced because cations are first released near the wells by proton-mediated mineral dissolution (as a result of the solubility trapping of $CO_2$) and then consumed by carbon mineralization away from the injecting point (Ellis and Peters, 2015; Fredd and Scott Fogler, 1998; Pokrovsky et al., 2005; Pokrovsky et al., 2009). Although this wave does not result in net consumption of $CO_2$ as is dictated by the stoichiometry of carbonation, it has a significant impact on the mechanical properties of a reservoir because it constantly relocates solid materials away from the injection wells. At the same time, hydraulic pressure decreases as fluid flow overcomes resistance from porous media, resulting in the degassing of $CO_2$. The degassed $CO_2$ forms dispersed bubbles which can be trapped in the very fine pores of chalk by capillary force. All these processes are determined by geochemically induced microstructural changes and are poorly understood in chalk formations.

We aim to fill the aforementioned information gaps in this study by combining an analysis of cumulative surface with *in situ* X-ray imaging to study the dissolution of natural chalk samples in an imposed flow field under elevated $CO_2$ pressure. Cumulative surface is a conceptual tool devised to calculate the reactive volume of a porous medium and to predict its dissolution pattern (Yang et al., 2016a, b; Yang et al., 2016c). It helps us identify the morphological features stemming from chemical reactions and hence distinguish between geochemical and geomechanical impacts on structural evolution. *In situ* X-ray tomography records the disintegration of chalk samples in real time with high fidelity and thus provides direct evidences of the complex coupling between processes.

2. **Materials and methods**

Synthetic calcite particles and ground chalk samples collected from the drill cuttings from the North Sea Basin were used for cyclic powder dissolution experiments. A detailed analysis of the samples' surface chemistry can be found in Okhrimenko et al (Okhrimenko et al., 2014). The synthetic calcite was purchased from Merck (Suprapur®) and had a grain size of ~10 μm. Each dissolution cycle had 3 runs where the solid remaining of the previous run was used as the starting material of the following run (Table S1). The starting mass of both the synthetic calcite and the natural chalk were 7.0 g for the first runs, 4.7 g for the second runs and 2.0 g for the third runs. 600 ml MilliQ water pre-equilibrated with 1 bar of $CO_2$ under room temperature (~25 $^o$C) was used as the starting solution in each run. Specific surface area ($m^2/g$) before and after dissolution was measured using the BET method. The experiments were conducted in closed batch reactors with magnetic stirring (310 rpm). Samples of the solution (0.1 mL) were taken with pipets, filtered (polypropylene, 0.22 μm pore size), diluted with 9.9 mL solution of 2% $HNO_3$ and 0.1% KCl and analyzed with atomic absorption spectroscopy (AAS). The pH of the solution was recorded before acidification.

Percolation experiments with continuous X-ray imaging used samples from a chalk outcrop near Aalborg, Denmark (Rørdal Quarry, Maastrichtian age). The samples were predominantly $CaCO_3$ coccoliths skeletal debris with an average porosity of ~45% and permeability in the range of 3-5 mD. Silica content of the samples was estimated to be ~ 4% and the specific surface area measured with B.E.T. was 7.31 $m^2/g$ (Okhrimenko et al., 2014). The chalk samples were machined into cylinders of 900 μm diameter and ~2 mm length, wrapped with a thin layer of heat shrinking polymer (3M, Maplewood, MN, USA) and placed between two stainless steel needles. The needles were hollow, 28 mm long and partially inserted into the polymer wrapping in alignment with the cylinder axis. The junctions between the polymer and the needles were sealed with epoxy. The total length of the composite sample, including the chalk cylinder, the polymer and the needles, was 58 – 60 mm. The composite sample was then loaded into a miniature of a Hassler core holder (Fig. S1) (Yang et al., 2016c) and confined with MilliQ water. An HPLC pump (Scientific Instrument, Series II) was used to compress the fluid. The confining side of the cell was sealed once the pressure of the fluid reached 10 bar.

Figure S2 shows the schematic setup for *in situ* X-ray imaging during the percolation experiments. MilliQ water was pre-equilibrated with a specified partial pressure of $CO_2$ under elevated temperature (1 bar at 25 $^o$C and 8 bar at 50 $^o$C) in a 600 ml compact laboratory reactor made of alloy C-276 (Series 5500 HP, Parr Instrument Company). The reactor is equipped with a stirrer, a gage, a heater and a temperature controller. Communication software on a connected laptop recorded the pressure and temperature history of the vessel. The partial pressure of $CO_2$ was modulated by a syringe pump (Teledyne ISCO 260D). Polyether ether ketone (PEEK) tubes were used to connect all accessories to the rotatory sample holder. During scanning all components were outside the radiation zone except the high pressure cell, which was mounted vertically on the rotation stage. The $CO_2$ - saturated fluid was injected axially from top to bottom.

*In situ* X-ray microtomography was conducted at the imaging beamline P05 of PETRA III at the Deutsches Elektronen-Synchrotron (DESY), operated by Helmholtz-Zentrum Geesthacht (Wilde et al., 2016). The dissolution process was monitored by recording consecutive tomographic datasets of 1200 projections over 180° at 28 keV and 1050 ms exposure. The time resolution was approximately 110 min/scan. The projections of 3056 by 3056 pixels were acquired with a 20× objective lens yielding an effective two-times binned pixel size of 1.33 μm. All projection data were down-sampled by a factor of two and converted to normalized sinograms by individual correlation of projections and flat field images that were recorded every 100 projections. In the sinograms we tracked the signal of the aluminum wall interface and matched it with a reference sinogram of the flow cell to compensate for any positional changes and distortions that may degrade the reconstructions. Such positional changes are caused by the necessity of taking frequent flat field images at the microtomography beamline at PETRA III. Distortions in the sinograms result from the tubing connected to the flow cell. When positioned poorly the tubing exerts a continuously changing dragging force on the flow cell during rotation which prohibits identifying a uniform center of rotation for the reconstruction of that timestep. Potential ring artifacts were suppressed by applying a Fourier-wavelet based destriping filter by Münch et al. (Gürsoy et al., 2014) to the sinograms and reconstruction was performed after manually identifying the center of rotation using the GridRec reconstruction algorithm (Dowd et al., 1999; Rivers, 2012). Both algorithms are packaged with TomoPy 1.0.0 (Gürsoy et al., 2014). Noise and remaining artifacts in the reconstructions were treated by applying four iterations of iterative non-local means denoising as described by Bruns et al. (Bruns et al., 2016b) using the same noise level estimate for all reconstructed timesteps. Finally, the timesteps were aligned spatially by digital volume correlation using Pearson correlation coefficient as a quality metric.

### 3. Results and discussion

#### 3.1 Powdered chalk dissolution in a closed free drifting system

The chalk dissolution rate depends on the aqueous composition. The most important parameters are the pH of the solution and the chemical affinity of the dissolution reaction (Lasaga, 2014). There can also be unidentified compounds released from natural chalk that modify the kinetics. In the segregated flow model explained in Section 3.2, the rate of chalk dissolution in a closed free drifting system is essential for predicting the reactive volume of intact chalk samples in an imposed flow field. "Free drifting" means the composition of a solution is not controlled and new equilibriums establish as substances are added (Brantley et al., 2008). This form of kinetics data relates the reaction rate to the extent of dissolution, i.e. the total amount of chalk dissolved in a batch reactor, and reflects how the rate decreases as the reaction approaches equilibrium. The extent of dissolution, measured in terms of the aqueous $Ca^{2+}$ concentration, is important because it changes the pH and the chemical affinity simultaneously, and may also determine the amount of rate modifying compounds released (Hassenkam et al., 2009). We chose to conduct our own kinetics measurements because 1) the rate law of chalk dissolution, especially one that describes the dependence of the dissolution rate on chemical affinity, is not available; 2) Natural chalk may contain rate modifying compounds whose impact on dissolution can only be examined experimentally. The aim of the data processing is to delineate the effect of the changing aqueous composition from that of the decreasing reactive surface area (Yang et al., 2014a), and to obtain the rate of chalk dissolution as a function of dissolved Ca concentration.

At any given instant, the aqueous concentration of $Ca^{2+}$ ($C_A$, mol/m$^3$) in a batch reactor is governed by

$$V \frac{dC_A}{dt} = \frac{N_A}{\tau_A}, \quad (1)$$

where $V$ represents the volume of the solution (m$^3$), $t$ the dissolution time (s), $N_A$ the amount of reactive sites on the mineral surface (mol) and $\tau_A$ the characteristic time scale (s) of calcium release. $\tau_A$ is proportional to the reciprocal of the intrinsic reaction rate and depends only on the composition of the solution. $N_A$ is time dependent and reflects the dynamics between the removal of reactive sites by dissolution and the emergence of internal sites by surface renewal (Yang et al., 2014a; Yang et al., 2014b). Hence, we look for a solution in the form $N_A = N_{A,ss} + N_A(t)$, where $N_{A,ss}$ represents a steady state solution and $N_A(t)$ is a transient term that describes the reduction of available reactive sites during dissolution. Let $N_{A,ss} = \chi_\infty N_{A0}$, where $N_{A0}$ is $N_A$ when $t = 0$ and $\chi_\infty$ the percentage of renewable $N_A$ as $t \rightarrow \infty$. The governing equation for $N_A$ is therefore

$$\frac{dN_A}{dt} = -\frac{N_A(t)}{\tau_A}, \quad (2)$$

which gives

$$N_A = \chi_\infty N_{A0} + (1-\chi_\infty) N_{A0} \cdot e^{-\int_0^t \frac{dt}{\tau_A}}, \quad (3)$$

leading to

$$\tau_A \left(\frac{dC_A}{dt}\right) = \frac{\chi_\infty N_{A0}}{V} + \frac{(1-\chi_\infty) N_{A0}}{V} \cdot e^{-\int_0^t \frac{dt}{\tau_A}}. \quad (4)$$

This integral differential equation regarding $\tau_A$ is a Fredholm equation of the second type (Pogorzelski, 1966). It can be solved by noticing that, with sufficiently small time intervals, the exponential term on the right hand side can be written as

$$\exp\left(-\int_0^{t_1} \frac{dt}{\tau_A(t)}\right)\exp\left(-\int_{t_1}^{t_2} \frac{dt}{\tau_A(t)}\right)\cdots\exp\left(-\int_{t_{n-1}}^{t} \frac{dt}{\tau_A(t)}\right) \approx \left(1-\frac{\Delta t_1}{\tau_{A,1}}\right)\left(1-\frac{\Delta t_2}{\tau_{A,2}}\right)\cdots\left(1-\frac{\Delta t_n}{\tau_{A,n}}\right), \quad (5)$$

where $n$ is the total number of data points. Back substitution gives a system of $n$ quadratic equations that can be solved recursively

$$\left.\frac{dC_A}{dt}\right|_{t_1} \cdot \tau_A^2(t_1) - \frac{N_{A0}}{V}\tau_A(t_1) + (1-\chi_\infty)\frac{N_{A0}}{V}\Delta t_1 = 0$$

$$\left.\frac{dC_A}{dt}\right|_{t_2} \cdot \tau_A^2(t_2) - \left[\chi_\infty\frac{N_{A0}}{V} + (1-\chi_\infty)\frac{N_{A0}}{V}\left(1-\frac{\Delta t_1}{\tau_A(t_1)}\right)\right] \cdot \tau_A(t_2) + (1-\chi_\infty)\frac{N_{A0}}{V}\left(1-\frac{\Delta t_1}{\tau_A(t_1)}\right)\Delta t_2 = 0 \quad (6)$$

...

$$\left.\frac{dC_A}{dt}\right|_{t_i} \cdot \tau_A^2(t_i) - \left[\chi_\infty\frac{N_{A0}}{V} + (1-\chi_\infty)\frac{N_{A0}}{V}\cdot\prod_{k=1}^{i-1}\left(1-\frac{\Delta t_k}{\tau_A(t_k)}\right)\right] \cdot \tau_A(t_i) + (1-\chi_\infty)\frac{N_{A0}}{V}\cdot\prod_{k=1}^{i-1}\left(1-\frac{\Delta t_k}{\tau_A(t_k)}\right)\cdot\Delta t_i = 0$$

Only the greater conjugate root (+) is physically realistic. We have

$$\tau_A(t \to 0) = \frac{H + \sqrt{H^2 - 4\cdot(1-\chi_\infty)H\cdot\Delta t_1 \cdot \left(\left.\frac{dC_A}{dt}\right|_{t=0}\right)}}{2\cdot\left(\left.\frac{dC_A}{dt}\right|_{t=0}\right)}, \quad (7)$$

where $H = SLR \cdot SSA \cdot \sigma_A$ is determined experimentally. *SLR* represents the solid to liquid ratio (g/m$^3$), *SSA* the specific surface area (m$^2$/g) and $\sigma_A$ the density of surface sites (mol/m$^2$). Similarly,

$$\tau_A(t_i) = \frac{\chi_\infty H + H_i + \sqrt{(\chi_\infty H + H_i)^2 - 4 \cdot H_i \cdot \Delta t \cdot \left(\left.\frac{dC_A}{dt}\right|_{t_i}\right)}}{2 \cdot \left(\left.\frac{dC_A}{dt}\right|_{t_i}\right)}, \quad (8)$$

where

$$H_i = (1 - \chi_\infty) H \cdot \prod_{k=1}^{i-1} \left(1 - \frac{\Delta t_k}{\tau_A(t_k)}\right). \quad (9)$$

Once $\tau_A(t_i)$ and $N_A = V(\chi_\infty H + H_i)$ are obtained by monitoring the concentration variation $\left(\left.\frac{dC_A}{dt}\right|_{t_i}\right)$ in a batch reactor, the chalk dissolution rate can be written as a function of total dissolved calcium ($C_A$) by noticing the monotonic mapping between $C_A$ and the elapsed time ($t$)

$$C_A(t) = \lim_{\Delta t \to 0} \left\{ \int_0^t \frac{SLR \cdot SSA \cdot \sigma_A}{\tau_A(t)} \left[ \chi_\infty + (1 - \chi_\infty) \cdot \prod_{k=1}^{i-1} \left(1 - \frac{\Delta t_k}{\tau_A(t_k)}\right) \right] \cdot d\dot{t} \right\}, \quad (10)$$

where $\dot{t}$ is a dummy variable.

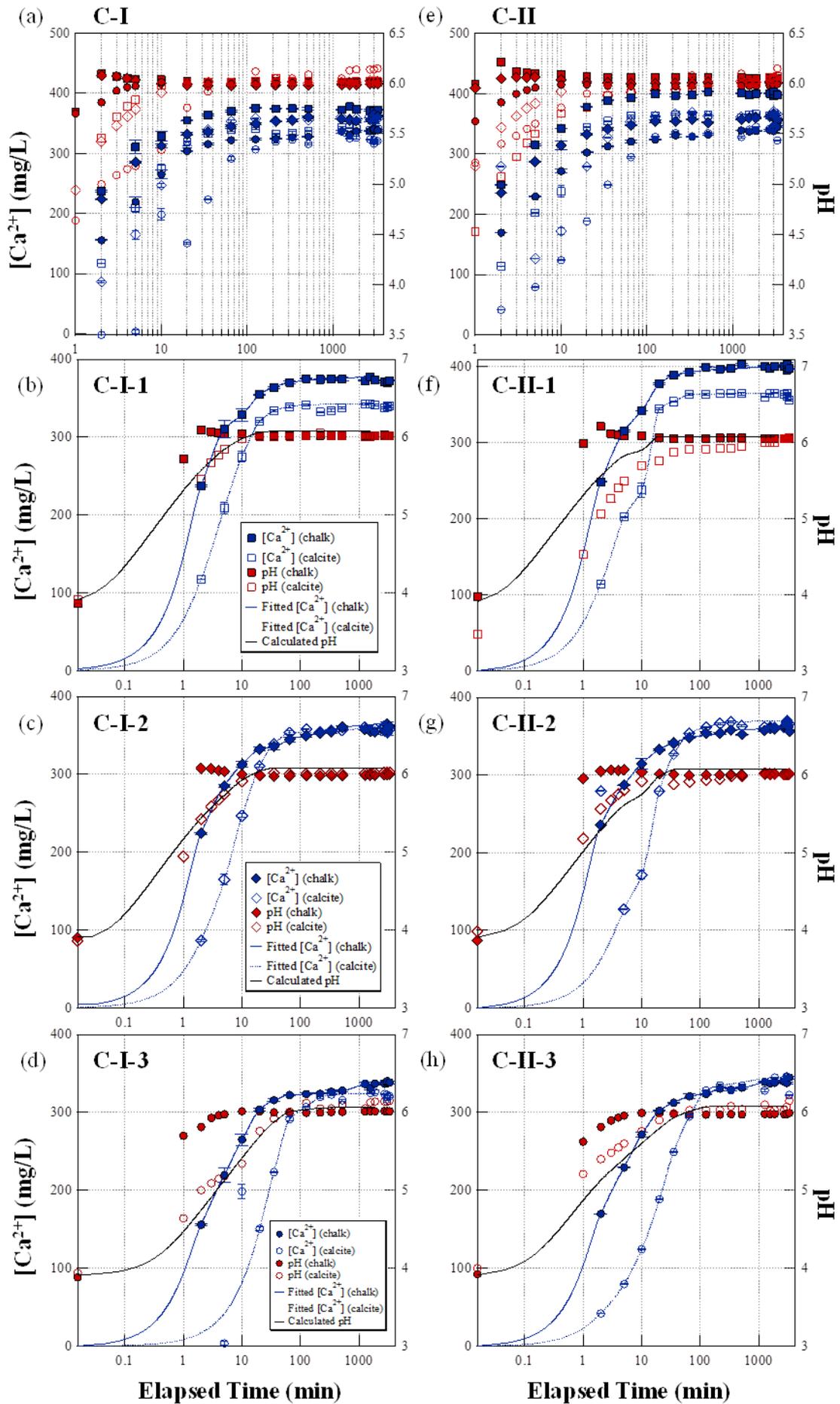

**Figure 1**. Cyclic dissolution experiments of chalk and synthetic calcite. (a) Cycle I data compilation. (b)–(d) Three consecutive runs of Cycle I. (e) Cycle II data compilation. (f)–(h) Three consecutive runs of Cycle II. The black solid lines show pH evolutions calculated assuming dissolution of pure calcite. The blue lines (solid and dashed) are calculated based on experimental data using equation (10).

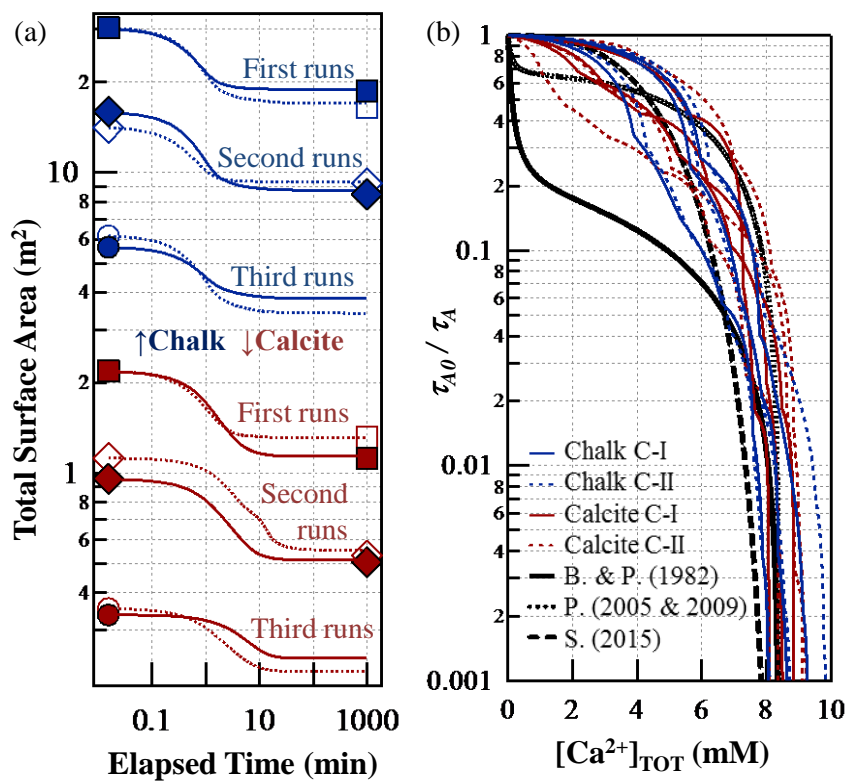

**Figure 2**. (a) Evolution of surface areas calculated with equation (9). Filled symbols: C-I; empty: C-II. (b) Decrease of dissolution rate as aqueous $Ca^{2+}$ concentration increases, reflected by an increase in the characteristic time $\tau_A$ calculated with equation (8), for the cyclic experiments. Also shown are curves based on rate laws proposed by Busenberg and Plummer (B & P) (Plummer and Busenberg, 1982), Pokrovsky et al. (P) (Pokrovsky et al., 2005; Pokrovsky et al., 2009) and Subhas (S) (Subhas et al., 2015).

Figure 1 compiles the results of two cyclic dissolution experiments (C-I and C-II). The experiments measured the evolution of $Ca^{2+}$ concentration as natural chalk and synthetic calcite dissolved in MilliQ water. Each cyclic experiment had three runs. The starting solid of the second and third runs were the remains of the previous runs. Performing a cyclic experiment has two advantages: it allows us to identify the chemical heterogeneities of natural chalk by comparing the results of different runs and circles; it also provides better constraints on the evolution of surface area (Figure 2a). Synthetic calcite served as a baseline for chalk reactivity. In both cycles, the greatest differences between chalk and calcite were observed in the first runs. Chalk showed a ~10.1% greater apparent solubility than calcite in C-I-1 (Fig.

1b) and ~10.4% in C-II-1 (Fig.1f). In the rest of the runs the two equilibrium concentrations were indistinguishable (Figs. 1c, 1d, 1g and 1h). Meanwhile, the pH approached the same equilibrium value (~6) in all runs. Compared to calcite the more rapid increase of [$Ca^{2+}$] for chalk in Fig. 1b-1h is attributable to the greater surface area (Fig. 2a), the solubility differences, under the same pH, suggested the impact of organics present as a result of biomineralisation.

Natural chalks have long been known to contain various organic compounds because of their biological origins (Bruns et al., 2016a; Hardman, 1982; Hassenkam et al., 2009; Okhrimenko et al., 2014). For example, elaborate coccospheres consisting of interlocked calcium carbonate platelets, accumulated during the sedimentary process and were covered by polysaccharides that controlled their formation. Other functional groups have also been identified on chalk surfaces by X-ray photoelectron spectroscopy (Bovet et al., 2015; Okhrimenko et al., 2013), and the inclusion of nanometer-thick organic layers leading to increased lattice strains in calcite structures have been analysed by X-ray diffraction (Poulsen et al., 2014). The soluble fraction of these compounds, once hydrolysed, may form complexation with aqueous cations and therefore increase a mineral's apparent solubility. A related but perhaps less plausible argument is that the apparent solubility of chalk was controlled by a more soluble and less stable form of $CaCO_3$. This was less likely given the long history of natural chalk under hydrothermal conditions. One would also argue that the phase controlling the solubility, if not transformed into calcite, should be carried on into the following experiment runs. However this possibility cannot be completely ruled out without a thorough understanding of how the presence of natural organics modifies the stability of mineral structures. Moreover, the very fine grain features of chalk (down to 2 nm) also pose the question whether the routine filtering (with 0.22 μm pores) and acidification operations needed for solution analysis are effective in isolating the dissolution products. The nanosized chalk particles may have penetrated the filters and therefore contributed to the measured $Ca^{2+}$ concentrations in the first runs of each circle. Despite the above mentioned uncertainties, Eq. 10 gives a fair description of concentration evolution based on the experimental data (blue lines in Fig. 1). Because the dissolution rate decreases dramatically (less than 1% of the rate far from equilibrium, Fig. 2b) when the Ca concentration reaches the uncertain regime (8.0 to 10.0 mM), the effect of the 10% difference in the apparent solubility on the reactive volume (discussed in Section 3.2) is negligible and is therefore not further discussed.

Figure 2 shows the delineated effects of surface area and solution composition. The symbols in Fig. 2a show measured BET surfaces before and after each run while the lines are computed from Eq. 9. The surface area displayed a net decrease over time as dissolution removed reactive sites and disintegrated fine particles (modelled by Eq. 2). Chalk showed an order of magnitude greater surface area than calcite, resulting in a rapid initial increase in Ca concentration and pH (Fig. 1). Figure 2b shows the dependence of $Ca^{2+}$ release rate on the extent of dissolution (Eqs. 8 and 9). The different runs of the same circle are plotted with the same type of line to show the variability of the kinetics within the same sample. Also shown are the dependences calculated based on Busenberg and Plummer (B & P) (Plummer and Busenberg, 1982), Pokrovsky (P) (Pokrovsky et al., 2005; Pokrovsky et al., 2009) and Subhas (S) (Subhas et al., 2015). The former two combined surface complexation models (SCM) with a sigmoidal Gibbs energy dependence based on the transition state theory (TST), while Subhas proposed an alternative mathematical form with a fractional power dependence. The formulation in this study (Eqs. 1-10) produced composition dependences that are, in general, less sensitive to the total dissolved $Ca^{2+}$ compared to the prediction by Busenberg and Plummer, especially when it is far from equilibrium. An important feature of the curves is their convexity. When the function is concave, reaction favours mixing and thus diffusion enhances dissolution, and vice versa (Danckwerts, 1958; Zwietering, 1959). The curve predicted by Pokrovsky et al. shows an inflection point near ~2 mM while the one by Subhas does not show any change of convexity. The curves computed from the experimental data show inflexions when the discretisation by Eq. 5 is affected by insufficient sampling frequency. If these discontinuities can be removed, we expect that the chalk dissolution curves were free of inflections and resemble the Subhas curve. On the other hand, the calcite curves are closer to the Pokrovsky curve. Overall, we conclude that (i) natural chalk contains contents, most likely organics of biological origins, that may change its apparent solubility in a closed free drifting system, although this has only a secondary effect on the reactive volume; (ii) the significantly greater surface area of chalk has a first order impact on the differences between chalk and calcite dissolution kinetics; (iii) the dependence of chalk dissolution rate on the dissolution extent does not deviate significantly from that of calcite. This dependence is, however, less sensitive to the increase of calcium concentration as is predicted by the SCM-TST based rate laws (Pokrovsky et al., 2005; Pokrovsky et al., 2009).

## 3.2  Segregated flow, reactive volume and dissolution pattern

In this section we explain why chemically induced pore growth, i.e., wormholing, is not expected with the given experimental sample size. Ruling out this possibility is important for arguing that mechanical property changes have contributed to the morphological evolution discussed in Section 3.3. We have previously shown that a prerequisite for wormholing in a dissolving porous medium is rate heterogeneity (Yang et al., 2016a, b; Yang et al., 2016c). This prerequisite means that the combination of flow field, solution composition and microstructure produces a reactant distribution that leads to a vastly differing mineral dissolution rate in space. In addition, this spatial distribution has to be sustainable before the depletion of solid. With a given sample size or simulation domain, this prerequisite can be examined by calculating the theoretical reactive volume. Reactive volume is the space encompassed by the dissolution front, whereas the latter is an isosurface on which the dissolution rate drops to zero along the flow direction. If the reactive volume is much smaller than the sample size, wormholes grow in an imposed flow field (Steefel and Lasaga, 1990, 1994). If the reactive volume is close to or greater than the sample size, wormholes may grow depending on how sensitive the mineral dissolution rate is to the extent of dissolution. A greater sensitivity leads to a greater likelihood of wormholing. Here we apply the segregated flow model (SFM) on a digital model of chalk to calculate the theoretical reactive volume. The SFM assumes a minimal extent of mixing within a flow field and therefore predicts the maximal spatial variations of reactivity in a simulation domain (Fogler, 2016). We show that the experimental sample size is similar to the theoretical reactive volume. With the injecting solution composition, the sensitivity of the chalk dissolution rate to the regime of dissolution extent within the sample size does not produce significant spatial variations of reactivity before solid depletion, even with the overestimated reactivity heterogeneities predicted by the SFM. We therefore expected a homogeneous dissolution pattern without wormholing. Our prediction was confirmed by *in situ* X-ray microtomographic evidences discussed in Section 3.3.

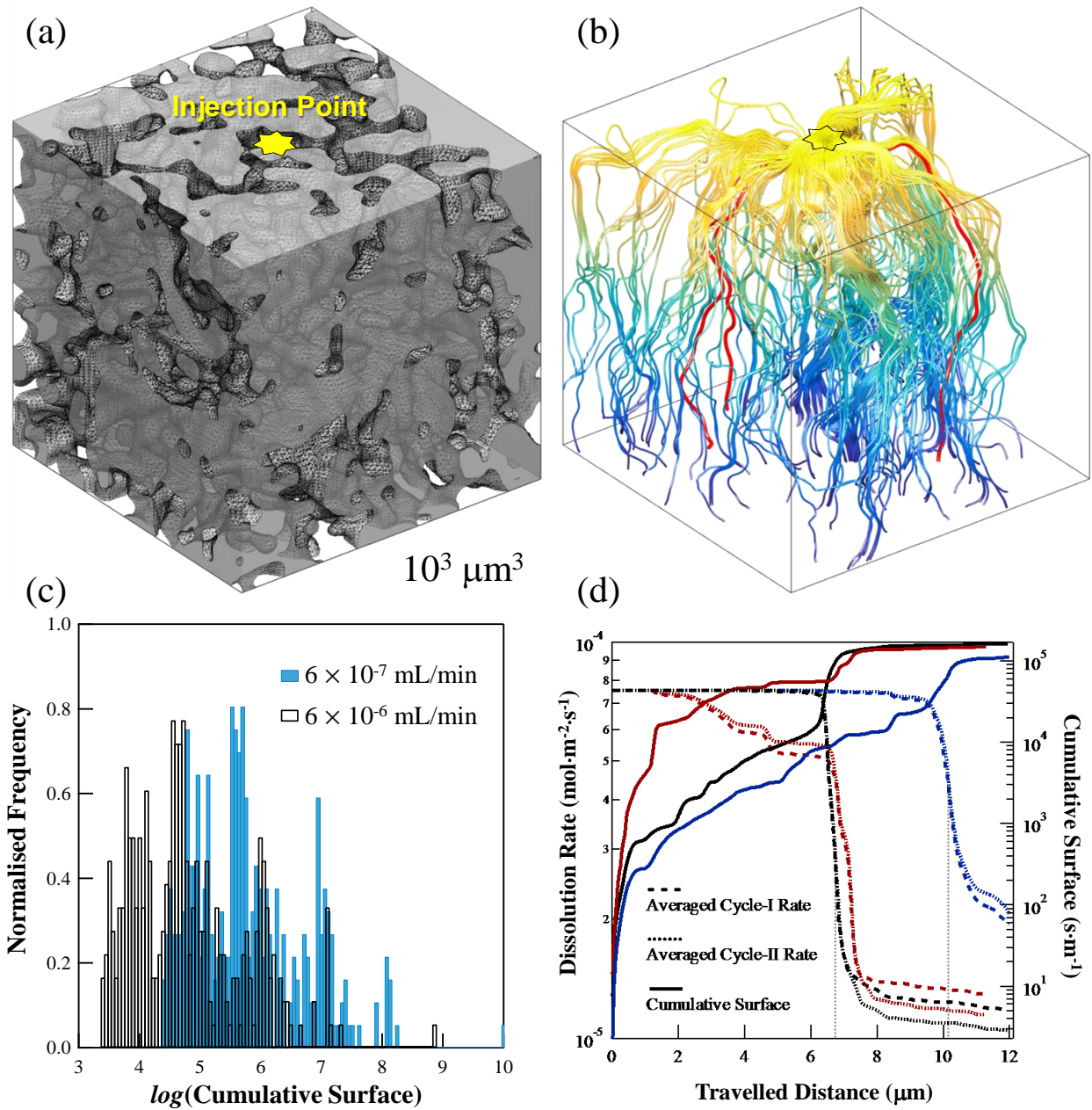

**Figure 3**. (a) A digital model of natural chalk microstructure obtained by x-ray nanotomography. (b) Streamlines originating from the injection point of fluid in a steady state flow field. The three red streamlines are arbitrary chosen whose cumulative surfaces are plotted in (d). (c) The distributions of cumulative surface of streamlines at two injection rates. The distributions are computed based on 400 streamlines in each flow field. (d) The cumulative surfaces of streamlines as fluid parcels travel through them and the decrease of dissolution rate as the reaction approaches equilibrium and their averaged chalk dissolution rates from different experimental cycles are compared. The three colours correspond to different streamlines (red in b).

Figure 3 explains how the theoretical reactive volume can be evaluated using the segregated flow model (SFM). SFM considers all streamlines originating from the fluid injection point as plug flow reactors (PFRs) (Fogler, 2016). These reactors have infinitesimal cross sections and are separated by non-permeable membranes of zero thickness (Fig. 3b). Fluid entering the sample travels as a collection of parcels along different streamlines and gains access to surface areas in different regions. If the parcels do not exchange mass and energy ("segregated"), the concentration of dissolution product, $[Ca^{2+}]_{TOT}$, at any point of the sample can be related to the residence time of the fluid parcel travelling through the streamline that passes the specific point by the performance equation of a PFR

$$\int_0^\tau SSA \cdot dt = \int_0^{[Ca^{2+}]_{TOT}} \frac{dC_A}{r_A} \qquad (11)$$

Where $SSA$ is the specific surface area along a streamline (m$^{-1}$), $\tau$ is the residence time of the fluid parcel at the point in question (second) and $r_A$ is chalk dissolution rate (mol·m$^{-2}$·s$^{-1}$). SSA is highly spatial dependent. For each fluid parcel the time available to access the surface area at a given point depends on the local flow field. The left hand side of Eq. 11 is therefore a strong function of sample microstructure and is hereafter termed Cumulative Surface (CS, s·m$^{-1}$). Figure 3c shows the distributions of CS for the given microstructure (Fig. 3a) at two injecting flowrates. The fluid is introduced at the top center of the sample and removed by 10,000 sink terms distributed evenly on the bottom plane. Darcy flow is assumed in calculating the steady state flow field. The normalized frequency is computed based on 400 streamlines originating from the injection location. The values of CS scale with flowrate and may span a few orders of magnitude because of the inherent heterogeneities of natural chalk.

The right hand side of Eq. 11 reflects the contribution of chemical kinetics to reactive volume and can be evaluated by integrating the curves in Fig. 2b. It is valid when the calcium concentration can be used as a master variable for rate determination, i.e. there exists a monotonic functional mapping between the total dissolved calcium and every other factor that exerts an influence on dissolution rate. Examples of such factors considered in this study are the solution pH and the saturation index (SI), which, according to Fig.1, do show monotonic dependence on $[Ca^{2+}]_{TOT}$. Equation 11 can thus be used to relate the physical properties of a sample to the chemical properties of the fluid. In Fig. 3d we show the cumulative surfaces for three arbitrarily chosen streamlines (red-colored in Fig. 3b) and how the dissolution rate would decrease along them. The streamlines differ in lengths and are all slightly longer than 10 μm because of the sample tortuosity. Within the same flow field, cumulative surface does not scale linearly with travelled distance, or residence time, because of the regional variations of surface area. As a result, the position of the dissolution front differs for each streamline, leading to the complex geometry of reactive volume (Fig. 4). For example, the streamline represented by the blue curves in Fig. 3d is only 1.67% longer than the one represented by the black curves (12.20 vs 12.00 μm). However, if we consider the position of the sharp dissolution rate decrease the dissolution front (vertical grey dashed lines), the former has a 52.2% longer reactive length compared to the other (10.2 vs 6.7 μm). Fig. 3d also suggests that using the rate correlations from either C-I or C-II does not produce significant differences.

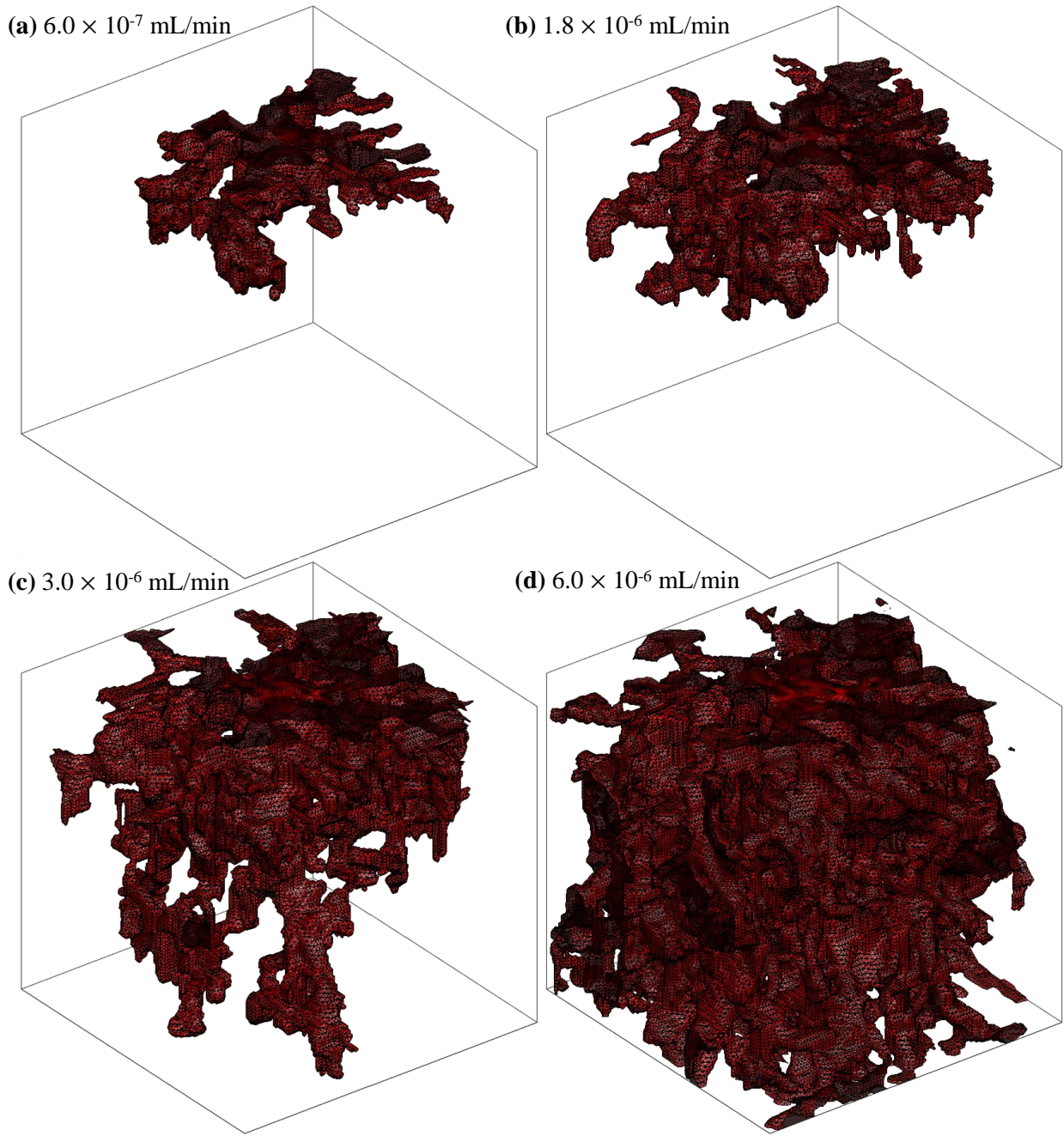

**Figure 4**. Reactive subvolumes of a sample given different injection rates. The sample geometry is shown in Fig. 3a. The isosurfaces are drawn at 10% of the reaction rate of the injecting fluid. The reactive subvolume can also be affected by the apparent solubility of chalk in solution. The relative size of the subvolume and the sample size determine the dissolution patterns recorded by an observer.

Figure 4 shows the effect of flowrate on the reactive volume within the same microstructure shown in Fig. 3a. The isosurfaces are drawn at $r_A = 7.56 \times 10^{-6}$ mol·m$^{-2}$·s$^{-1}$, 10% of the dissolution rate far from equilibrium at pH 3.91. The increase in flowrate expands the reactive volume by shortening the mean residence time and therefore providing more reactants. When the microstructure of a porous medium is modified only by geochemical reactions, the relative size of the region of interest (ROI) and the reactive volume determine the observable dissolution pattern. If the reactive volume is smaller than the ROI (e.g., Fig. 4a), the observer sees both the dissolving region and the non-reactive region in the ROI. As the microstructure evolves, the dissolving region channelizes the fluid and generates wormholes (often referred to as "wormholing" or "fluid focusing", Fig. 5b) ((Noiriel, 2015) and references therein). In contrast, if the ROI is of comparable size (e.g., Fig. 4d) or smaller than the reactive volume, the observer sees no significant difference in dissolution rate within the ROI. This is especially true when a buffering mechanism is involved in the presence of a weak acid like the carbonic acid and the rate is not sensitive to the release of $Ca^{2+}$. Examples of such include the Subhas rate law (Fig. 2b) and the chalk dissolution rate from this study (Fig. 6). A homogeneous dissolution pattern is therefore expected (Fig. 5c). In addition to changing the flowrate, the reactive volume can also be modified by changing the apparent solubility of the percolating fluid. For example, carbonated water equilibrated with 1 bar $CO_2$ dissolves 836 mg calcite/L and has an equilibrium pH of 6.08. A solution with hydrochloric acid (HCl) can have the same initial pH (3.91) but dissolves only 19 mg calcite/L and equilibrates at pH 9.41. As a consequence, the former solution provides orders of magnitude greater reactive volume than the latter given the same cumulative surface and is more likely to produce a homogeneous dissolution pattern.

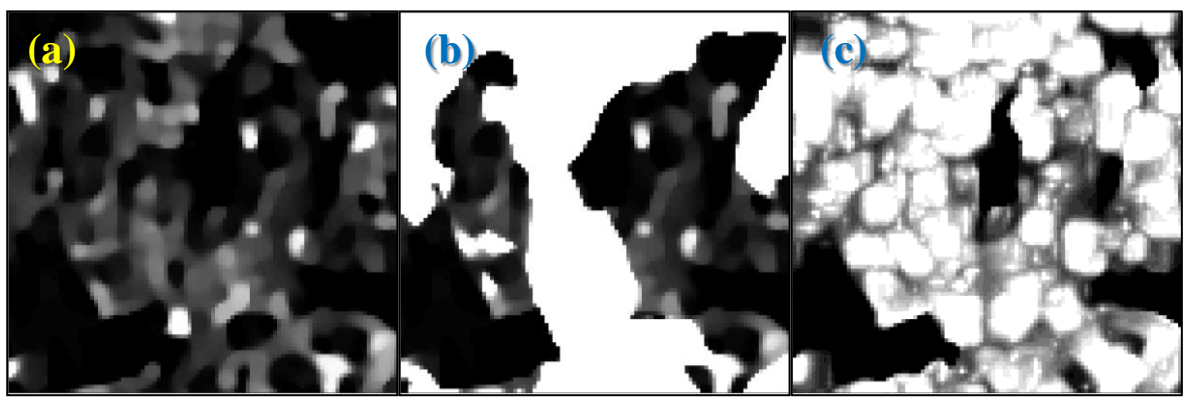

**Figure 5**. Dissolution patterns. (a) A cross section of the starting microstructure (porosity = 0.21). (b) A wormholing pattern. (c) A homogeneous dissolution pattern. Both (b) and (c) have an averaged porosity of 0.60. The graphs are produced by simulations based on Yang et al. (Yang et al., 2016a)

We conclude this section by emphasizing that, if the microstructure of the porous media is modified only by geochemical reactions, the observable dissolution patterns can be predicted by comparing the reactive volume and the ROI. The reactive volume can be calculated with the knowledge of the cumulative surface and the dissolution kinetics in a closed free drifting system. The ROI is typically the simulation domain in numerical experiments or certain portion of the field of view (FOV) in X-ray imaging.

### 3.3 Chalk percolation with *in situ* X-ray tomography

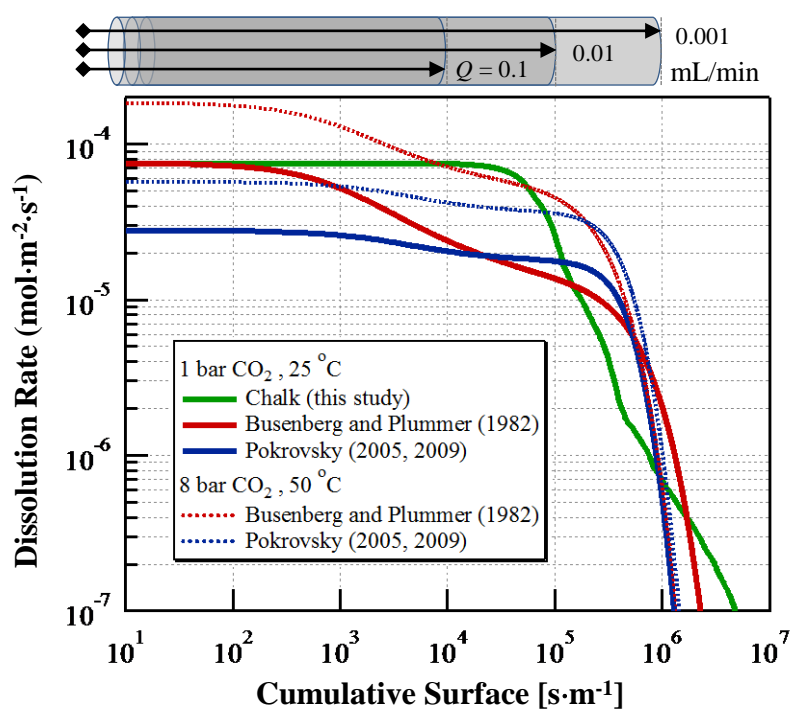

**Figure 6**. Decrease of $CaCO_3$ dissolution rate with increasing cumulative surface (CS) in plug flow. Solid lines show the results of this study as well as calculations based on published rate laws under ambient conditions. Dashed lines are estimated calcite dissolution rates in a solution pre-equilibrated with 8 bar $CO_2$ under 50 $^oC$ (the dissolution takes place under 25 $^oC$). Above the graph an estimate of the cumulative surface for a cylindrical chalk sample, as used in our experiment, can be found. It scales with the fluid injection rate.

The Aalborg chalk used in this tomography study has an average specific surface area of 7.31 $m^2/g$ (Okhrimenko et al., 2014), leading to an average cumulative surface (CS) on the level of $10^5$ s/m with a volumetric flowrate of 0.01 ml/min (Yang et al., 2016a, b; Yang et al., 2016c). This CS is comparable with that what is needed for a sharp rate decrease predicted by the different rate equations (Fig. 6). Decreasing the flowrate leads to an increase in the CS for a given sample (and vice versa) as is demonstrated on the top of Fig. 6. We therefore expect the sample to dissolve homogeneously with the given experimental setup. This pattern was indeed observed during the early stage of dissolution up to 24 hours. Figure 7 shows the evolution of the sample morphology recorded by *in situ* X-ray tomography over a period of 34 hours. The fluid was equilibrated with 1 bar of $CO_2$ under 25 °C and percolated downwards through the sample. The morphological changes during the first 16 hours of percolation were not significant, although the X-ray absorption of the complete FOV decreased gradually, manifested by a decrease in image brightness. At $t = 17.6$ hr the flowrate was increased to 0.04 ml/min to facilitate the dissolution. The increase of fluid velocity further decreased the residence time and thus the CS, making a greater portion of the sample dissolving at a far-from-equilibrium rate. Three distinct features of chalk dissolution were observed. First, the greyvalues over the whole FOV decreased concurrently, confirming that the dissolution took place at comparable rates both near and far from the fluid entrance. Second, the contrasts near the coccospheres and other fossil-shaped entities increased, signifying the presence of reactivity heterogeneities within micrometer scale. The preservation of the fossil shapes is related to the greater durability of $CaCO_3$ with biological origins. This reactivity difference is also in concord with the differences observed in the consecutive runs of the cyclic experiments (Section 3.1), which signifies the uncertainties associated with treating chalk as a sole phase. Third, starting from $t = 23.7$ hr, a tailing artefact started to appear within the sample (e.g., no longer spherical coccospheres), but neither in the polymer sleeve nor in the wall of the aluminum tube. This artefact is typical when the reconstruction center of the radiographs is poorly specified. However, the sharp reconstructions of the other entities in the FOV (polymer and Al tube) indicated that this was the result of the disintegration and motion of the sample during a scan. This motion of the disintegrated sample continued even after the flowrate was decreased to 0.02 ml/min at $t = 23.7$ hr and lasted until the end of data collection. The appearance of a large void space first near the fluid outlet ruled out the possibility of wormholing, which would always occur in the flow direction. We thus conclude that the morphological evolution observed from 25.6 hour to 33.4 hour took place because the mechanical property of the partially dissolved sample was not able to sustain the dragging force exerted by the flow field. As a result, the remnants moved in a concerted manner to form a microstructure that minimized the fluid resistance and therefore the dragging force.

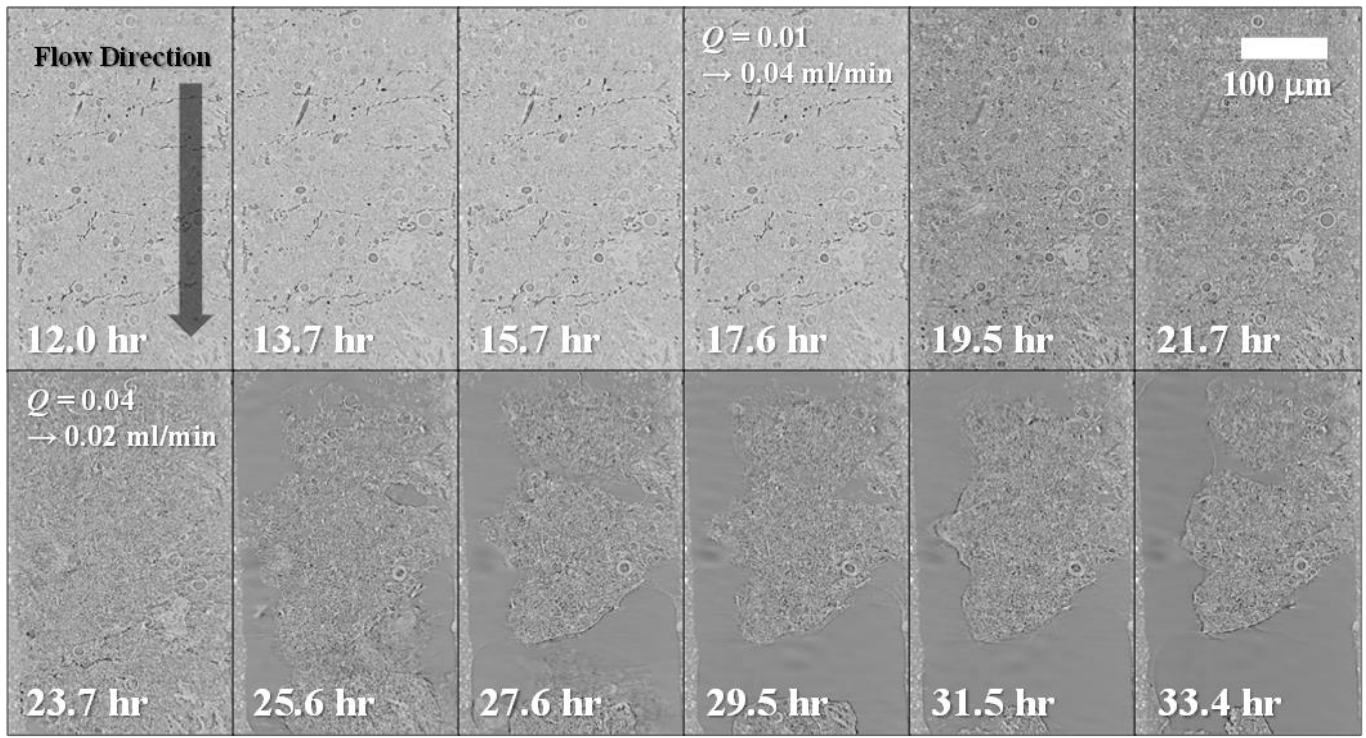

**Figure 7**. Microstructural evolution of a cylindrical chalk sample during a 33.4 hours percolation recorded by *in situ* X-ray tomography. The images are cross sections of the scanned volume along the axis of the sample. The percolating fluid was MilliQ water pre-equilibrated with 1 bar $CO_2$ under 25 °C flowing from top to bottom.

Figure 8 shows the morphological evolution of a chalk sample over an 86.6 hours percolation. The sample was flooded with MilliQ water pre-equilibrated with an elevated pressure of $CO_2$ under 50 °C. The pressure of the pre-equilibrium chamber was set to 5 bar for the first 20 hours and increased to 8 bar for the rest of the data collection (Figure S3). The fluid was driven through a 3 meter long PEEK tube before entering the sample holder and the dissolution took place inside the experimental hutch under ambient temperature (~25 °C). The flowrate was regulated by six pieces of plate-shaped fittings, three on each side inside the sample holder, to be approximately 0.016 ml/min on average despite the changes of the pressure drop across the holder. An additional safety threshold was set through the pump controller so that the flowrate would not exceed 0.02 ml/min at any time. The cumulative surfaces required to reach the dissolution front, based on the two TST-SCM rate laws, are plotted in Fig. 6. We expect the sample to dissolve homogeneously because of the comparable sizes of the reactive volume and the sample.

A homogeneous dissolution pattern was observed for up to 20 hours, after which we recorded three typical geomechanical effects that changed regional porosity in very different manners: compaction, fracturing and grain relocation. From $t = 7.7$ to 15.0 hr, the contrast near the coccospheres at the lower center of the FOV increased without substantial morphological change. This indicated that the dissolution took place both near and far from the fluid entrance. Starting from 16.8 hr, dissolution-generated porosity started to appear near the fluid entrance, but not yet farther downstream. This is consistent with the predictions made with the two rate laws in Fig. 6, which show both faster dissolution and a greater sensitivity to dissolution progress at lower CS ($< 10^{-3}$ s·m$^{-1}$) under elevated $CO_2$ pressure. From 20.4 hr onwards, the disintegrated structure near the fluid entrance started to get compacted by forces exerted by both the flow and the confining fluid outside the polymer sleeve. At 26.6 hr, a pronounced pore originated from the fluid entrance, against our anticipation of no wormhole. This pore initiated a successive growth of porous structures towards the fluid outlet. From 65.8 hr onward, accompanying the pore propagation, grains of various size were mobilized in the flow field by the dragging effect of the fluid. These observed mechanical effects are qualitatively different in changing the regional porosities and are discussed below.

The sign of feedback between mineral dissolution rate and microstructure permeability is essential in characterising a developing flow system in porous media. In general, this feedback is positive in an imposed flow field. Fast dissolution makes the rock more permeable by increasing the local porosity, which shortens the residence time of the fluid, drives the dissolution farther away from equilibrium and thus speeds up dissolution even more. This positive feedback leads to an infiltration instability that determines the formation and evolution of ubiquitous ramified flow networks (Ortoleva, 1994). There are, however, dampening mechanisms for this instability. For example, molecular diffusion tends to smear out the concentration gradient of reactants and therefore reduces the spatial variations in the dissolution rate (Yang et al., 2016a, b; Yang et al., 2016c). Similarly, solid compaction dampens this positive feedback because it decreases the porosity of the faster dissolving regions. Figure 9 shows the zoomed-in images of the near-entrance region where the solid material was being compressed during the early stage of dissolution. The arrows show the direction of the forces from the confining fluid (radial) and the convective flow (axial). The yellow skeleton marks the relative positions of four coccospheres that were located near the center of compression. The pixel brightness ($F(\rho)$) in these images reflects the absorption of X-ray and therefore shows a positive correlation with the electron density of the material. This correlation means $CaCO_3$ is brighter while the fluid is darker and the variations of the greyvalues reflect the changes in the porosity. Similarly, the geometric surface can be defined as the frequency and amplitude of the spatial variations of material density (Yeong and Torquato, 1998). Consequently, the absolute value of the intensity gradient (F(SA)) in the images is positively correlated to the geometric surface area resolvable with X-ray imaging. In Fig. 10 we plot the averaged $F(\rho)$ and F(SA) for the images shown in Figs. 8 and 9. The dissolution resulted in a continuous loss of solid, reflected by the monotonic decline of $F(\rho)$ averaged over the whole ROI. However, in the zoomed-in images, the porosity was sustained and momentarily increased because of the local compression. The evolution of surface area stems from a combined effect of two processes. Firstly, the compaction naturally led to the concentration of grain surfaces in the ROI. Secondly, the geometric surface area was amplified by the infiltration instability as a type of local heterogeneity (Yang et al., 2016b). This amplification can only be contained by a regional depletion of solid material. Hence, momentary increases of F(SA) were observed in the zoomed-in ROI as well as in the complete FOV. The variance bars are not associated with the expected error but reflect the heterogeneities inherent to chalk.

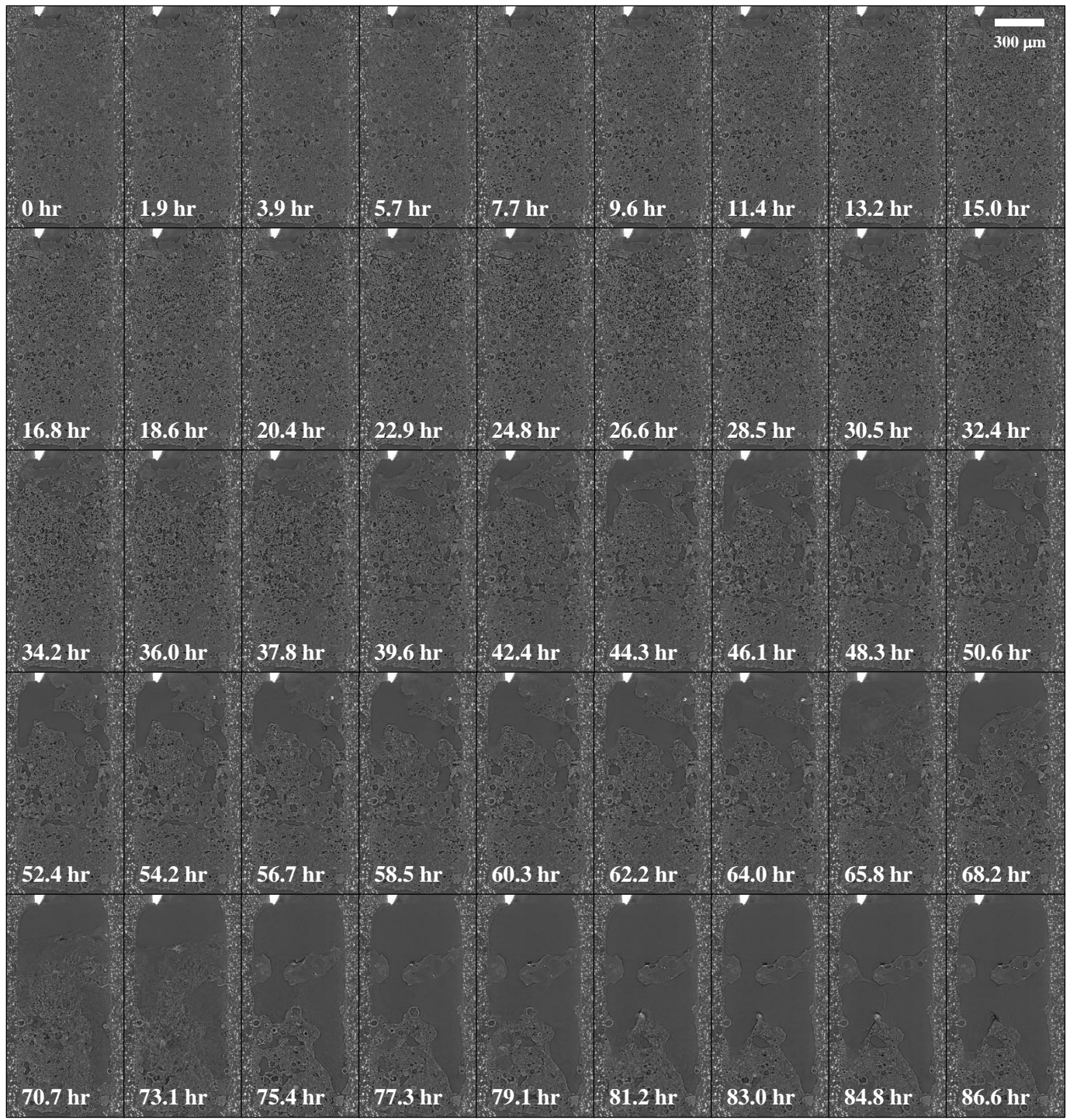

**Figure 8**. Microstructural evolution of a cylindrical chalk sample during an 86.6 hours percolation recorded by *in situ* X-ray tomography. The images are cross sections through the scanned volume passing through the axis of the sample. The percolating fluid was MilliQ water pre-equilibrated with ~8 bar $CO_2$ under 50 $^{o}$C flowing from top to bottom. The very bright spot on top is the stainless steel needle serving as the fluid entrance. The polymer wrapping on both sides of each image was not dissolved during the percolation.

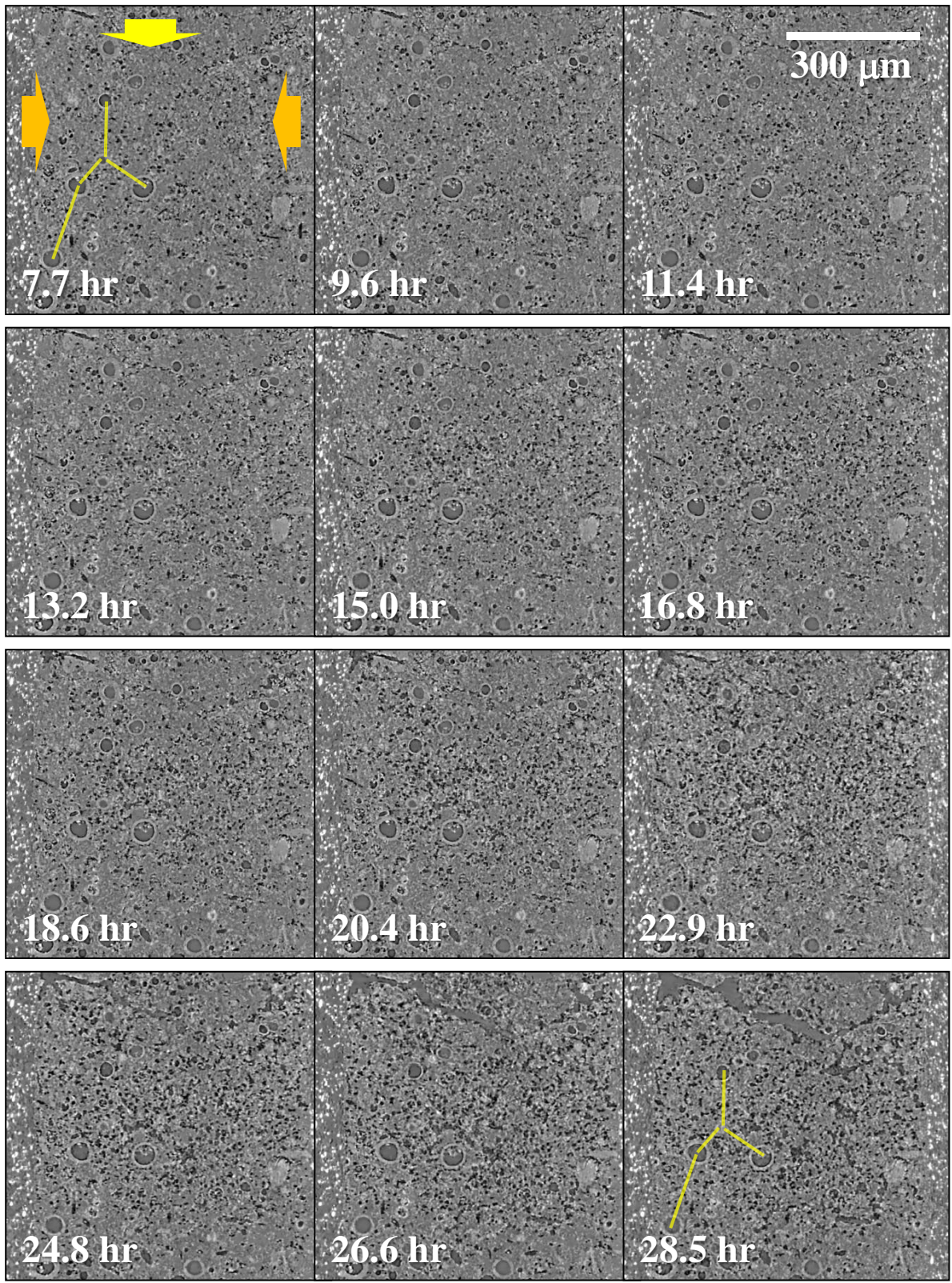

**Figure 9**. A zoomed in region near the fluid entrance showing the compaction of disintegrating materials during the initial stage of percolation. The yellow sticks mark the relative distances of the coccospheres that were close to the center of compaction. These biogenic calcium carbonates are covered by a trace amount of polysaccharide and are more resistant to dissolution.

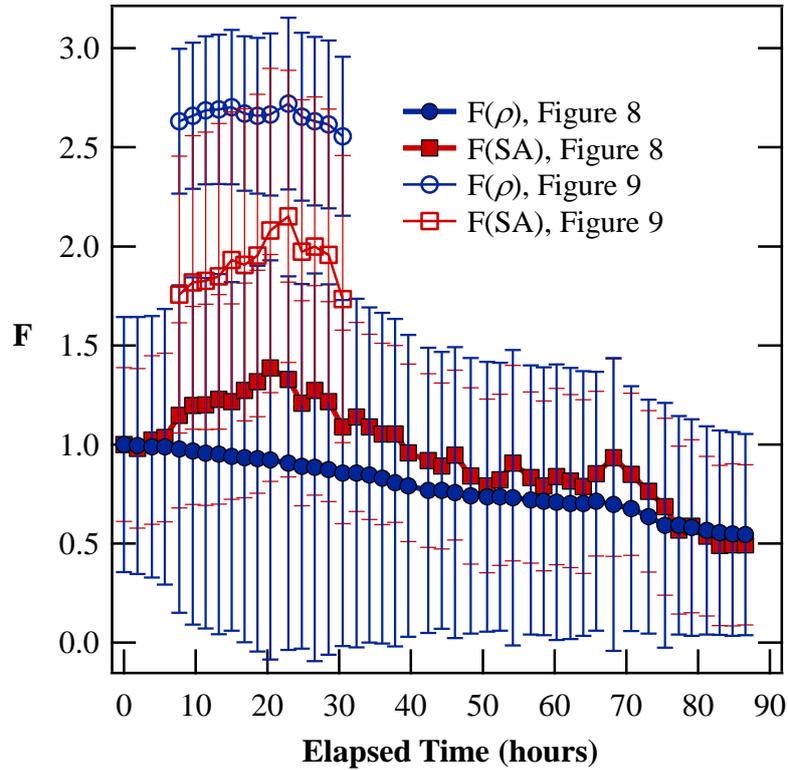

**Figure 10**. Evolution of averaged density, F($\rho$), and surface area, F(SA) based on figures 8 and 9. F($\rho$) is the normalized grey value of voxels that positively correlates with the sample density. F(SA) is the normalized absolute value of the gradient of F($\rho$) which positively correlates with the geometric surface area. The variance illustrates the internal heterogeneities of the sample.

The unexpected pore growth was a result of mechanical fracturing. Figure 11 shows the evidences that the preexisting structural defects were turned into pores as dissolution weakened the surrounding material. In contrast to wormholing, where the generated pores are always aligned with the flow direction, the fractures observed were perpendicular to the flow (yellow circles) and preserved the shapes of the original defects. Before the appearance of the first fracture, the amount of solid kept decreasing as was indicated by the F($\rho$) decline in Fig. 10. From $t = 30.5$ hr onward the pore started to grow from the fluid inlet. A greater area of the images later in the time series was then occupied by void space that ought to decrease F($\rho$) significantly. This change in the trend of F($\rho$) was nonetheless not observed in Fig. 10. We thus speculate that the expansion of the pore structure was not solely due to chalk dissolution but also to the fact that the materials were being pushed aside by hydraulic pressure. This mechanical fracturing constitutes an important positive feedback in the dissolution-induced pore growth because it increases the permeability of the faster dissolving regions while at the same time decreases that of the slower dissolving ones by pushing them towards less soluble grains.

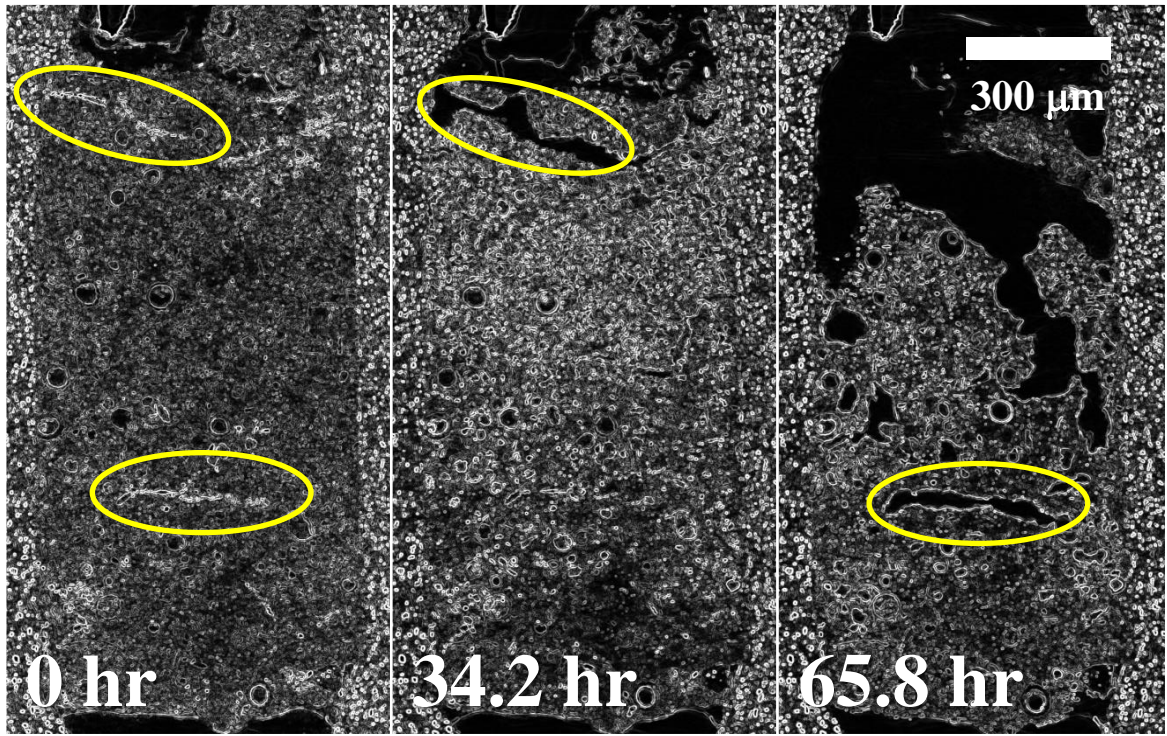

**Figure 11**. Preexisting structural defects (yellow circles at 0 hr) developed into fractures (yellow circled). The images show F(SA) instead of F($\rho$). The brighter pixels indicate qualitatively greater geometric surface area.

A third mechanical impact on the microstructural evolution was the mobilization and relocation of grains. This motion of undissolved material is typically associated with the dragging effect of the fluid flow and moves fine particles downstream. Figure 12 shows an example of grain relocation. At $t = 50.6$ hr a partially attached chalk grain near the fluid entrance (top yellow circle) started to move downwards. At $t = 62.2$ hr the particle moved slightly outside the viewing plane. At the same time the pore initiated by fracturing continued to grow, showing a greater degree of overall pore connectivity with widened pore necks. At $t = 70.7$ hr the detached particle landed on the less disintegrated structure below and the landing impact squeezed the previously discussed fracture (yellow ellipses). The motion of porous materials during a scan does not contribute significantly to the estimate of average porosity from X-ray intensity. This is because a point in space partially occupied by solid, either spatially or temporally, yields a grey voxel in the tomographic reconstruction representing the material density averaged over both the space of the voxel and the time of the scan. In contrast, grain motion may contribute to an overestimate of surface area based on the intensity gradients because of the multi-counting of the same surface of a moving grain at different locations during a single scan. This may have resulted in the momentary increase of F(SA) between 65 hr and 75 hr in Fig. 10. Overall, mobilization of less dissolvable solid has a strong dampening effect on channelized pore growth – it reduces the permeability of the dissolution front by both compaction and material relocation.

In summary, three types of geomechanical impacts on the microstructural evolution have been identified and their effects on the infiltration instability, a principal mechanism for pore growth in porous media, have been discussed. Fracturing enhances the instability while compaction and grain mobilization dampen the positive feedback between mineral dissolution and fluid flow, leading to a more stable dissolution front. In part II of this study we propose an explicit functional relationship between X-ray absorption and voxel porosity, based on which a comprehensive analysis of the 45 3D datasets collected over the 88 hours of percolation will be presented.

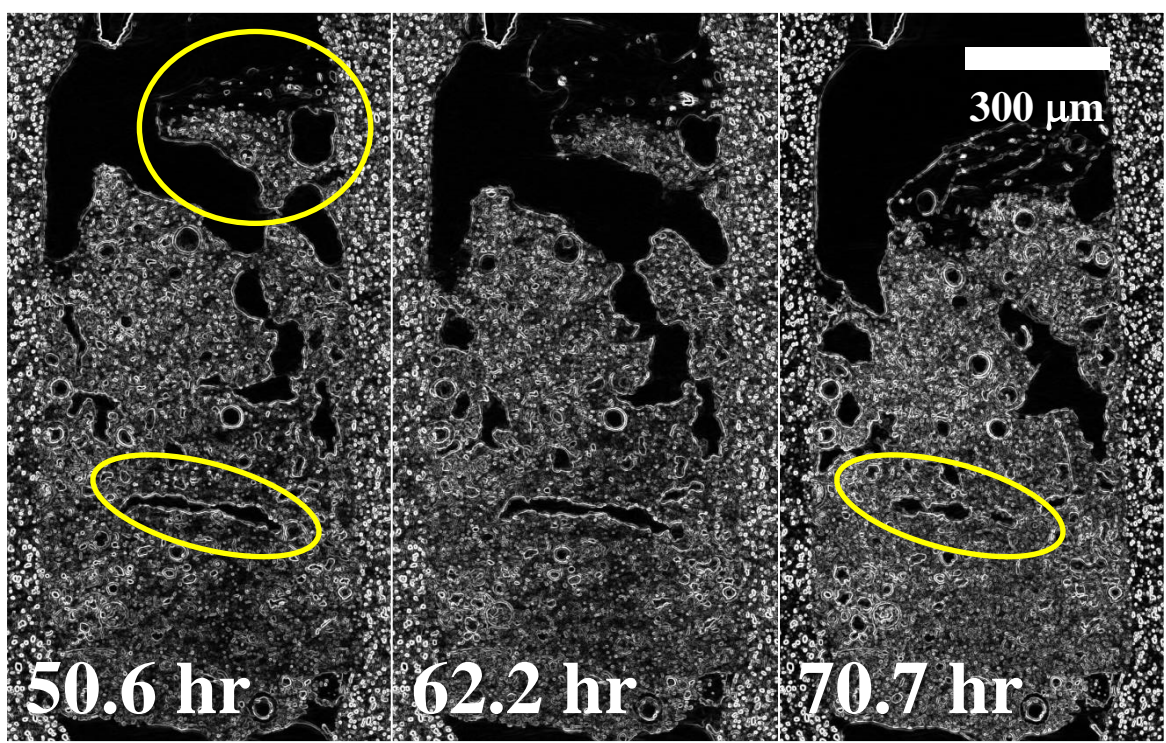

**Figure 12**. Relocation of a loosely attached grain (upper yellow circle at 50.6 hr) brings less dissolvable material to the reaction front. This concerted motion of particles may demonstrate additional mechanical impacts. For example, the fracture at 50.6 hr (lower yellow circle) was squeezed after the grain landed on top of the bulk sample, decreasing regional porosity. The images show F(SA) instead of F($\rho$). The brighter pixels indicate qualitatively greater geometric surface area.

## 4. Conclusions and implications

We showed that in a closed free drifting system chalk has a slightly higher apparent solubility than synthetic calcite. Chalk dissolution is less sensitive to the increase of aqueous $Ca^{2+}$ concentration compared to the predictions by published calcite dissolution rate laws. We demonstrated how to combine the kinetics data of a free drifting system with X-ray imaging to compute the reactive volumes in porous media with imposed flow fields. A comparison between the sample size and the reactive volume allowed us to predict the dissolution patterns. With *in situ* X-ray tomography, we showed that in a homogeneously dissolving chalk samples, geochemical reactions served as a trigger for geomechanical effects on microstructural evolution. Mechanically induced morphological changes were discussed in the context of infiltration instability. We concluded by stating that fracturing enhanced the positive feedbacks between mineral dissolution and fluid flow, while compaction and grain mobilization dampened the instability of dissolution front. We look for a definite functional relationship between the voxel level X-ray absorption and the porosity/surface area of reconstructed pore geometry, which will be the subject of the second part of this study.

The results revised our expectations for geologic structural evolution in carbon storage associated scenarios. We have previously shown that when a reservoir structure is modified only by geochemical reactions during supercritical $CO_2$ injection, wormholes form in short term because there exists no effective containing mechanism for reactive infiltration instability other than the precipitation of secondary minerals. The latter may entail for a long time before sufficient cations are scavenged. The effect of dissolved $CO_2$ is a reduction of the coupling strength between chemical reaction and mass transfer and it therefore increases the sizes of wormholes. The introduction of geomechanical effects hugely complicated this vision. Out of the three identified phenomena that are associated with an elevated $CO_2$ pressure, two of which provided local negative feedbacks on channelized pore growth. This postponement of fluid focusing may earn time for mineral precipitation and hinder the spontaneous development of underground fluid dissipation networks. Should this be the case, the energy consumption for $CO_2$ pumping shall increase, while the mechanical sustainability of the geologic formations near the injection wells may not deteriorate as expected. Nevertheless, the net effect of geochemical and geomechanical coupling is far from being well understood. The competition between wormholing/fracturing and compaction/grain relocation will have to be further examined.

**Acknowledgements**

We thank F. Saxlid for help with the design and manufacturing of the percolation cell and H. Suhonen at the ID22 beamline at ESRF (European Synchrotron Research Facility, France) for technical support. Parts of this research were carried out at the light source PETRA III at DESY (Deutsches Elektronen-Synchrotron, Germany. Beamtime grant ids I-20150242EC and I-20160208EC), a member of the Helmholtz Association (HGF). We thank Imke Greving and Fabian Wilde for help with data collection at beamline P05. Funding for this project was provided by the European Union's Horizon 2020 research and innovation programme under the Marie Sklodowska-Curie grant agreement No 653241, the Innovation Fund Denmark through the CINEMA project, as well as the Innovation Fund Denmark and Maersk Oil and Gas A/S through the P3 project. Support for synchrotron beamtime was received from the Danish Council for Independent Research via DANSCATT.

# Supplementary Information for

# Direct observation of coupled geochemical and geomechanical impacts on chalk microstructural evolution under elevated $CO_2$ pressure. Part I.

Y. Yang,[1]* S. S. Hakim,[1] S. Bruns,[1] M. Rogowska,[1] S. Boehnert,[1] J.U. Hammel,[2] S. L. S. Stipp[1] and H. O. Sørensen[1]

[1] Nano-Science Center, Department of Chemistry, University of Copenhagen, Universitetsparken 5, DK-2100 Copenhagen, Denmark

[2] Helmholtz-Zentrum Geesthacht, Max-Planck-Straße 1, 51502 Geesthacht, Germany

Current address: Universität Bremen, Bibliothekstraße 1, 28359 Bremen, Germany

Corresponding author: yiyang@nano.ku.dk

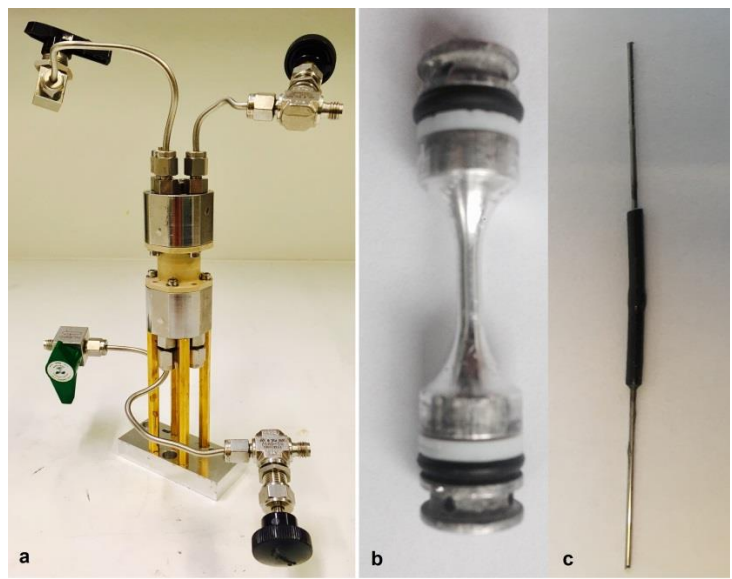

**Figure S1**. Photographs of (a) the miniature Hassler core holder used in *in situ* X-ray imaging, the aluminum central tube (b) and a composite sample (c). The chalk cylinder is wrapped in black heat shrinking tube.

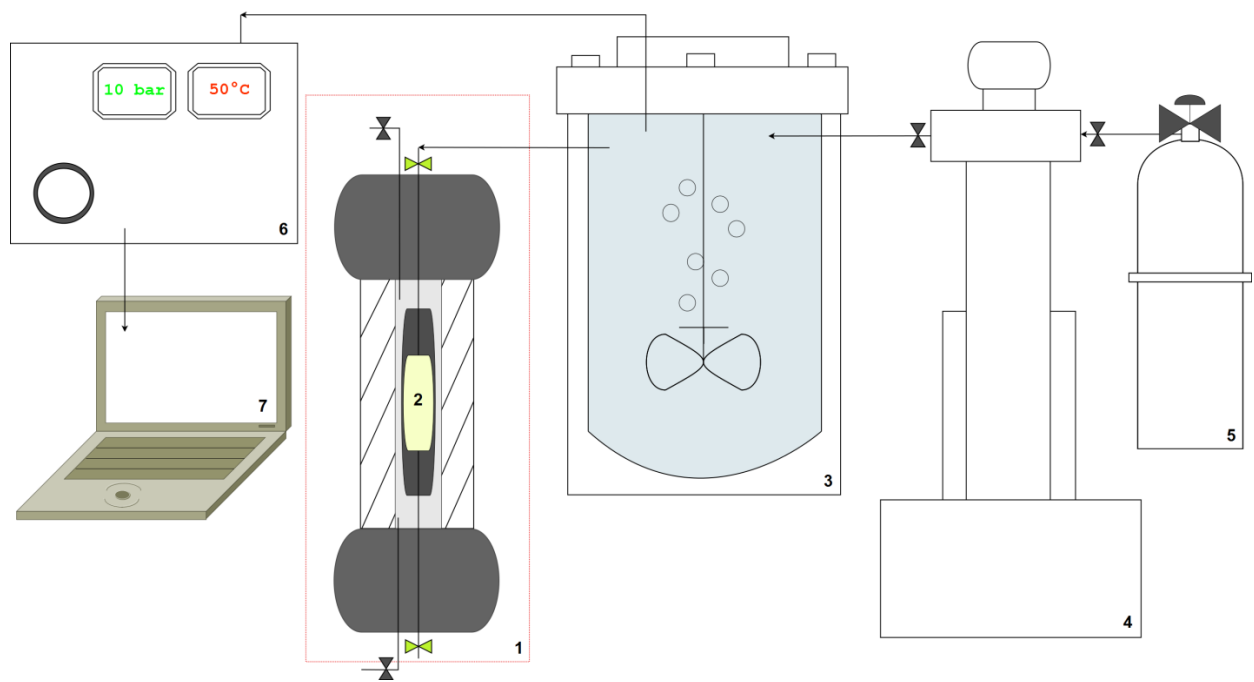

**Figure S2**. A schematic drawing of the *in situ* x-ray imaging experimental setup: 1) the core holder with 2) a chalk sample, 3) a high pressure reactor, 4) a syringe pump, 5) a $CO_2$ gas cylinder, 6) a pressure and temperature controller, 7) a recording device.

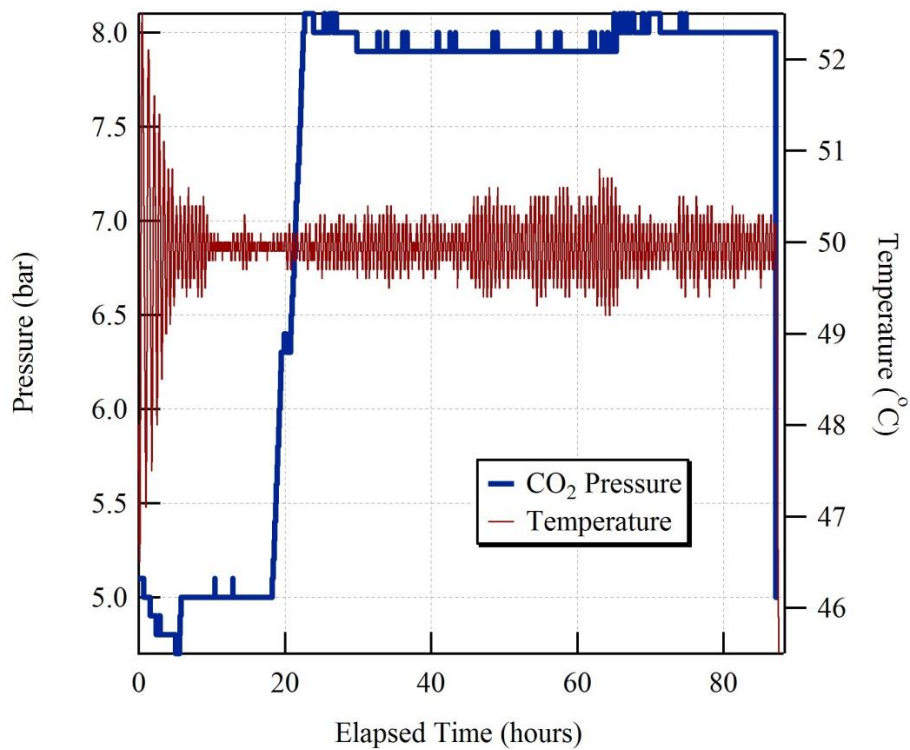

**Figure S3.** Pressure and temperature history of the pre-equilibrium vessel during the percolation experiment shown in Fig. 8

**Table S1.** Specific surface area during the cyclic dissolution experiments measured by B.E.T.

| B.E.T. SSA (m²/g) | Chalk | Synthetic Calcite |
|---|---|---|
| Original | 2.598 | 0.187 |
| Cyclic Experiment I run 1 | 2.036 | 0.122 |
| Cyclic Experiment I run 2 | 1.692 | 0.101 |
| Cyclic Experiment I run 3 | 2.269 | 0.145 |
| Cyclic Experiment II run 1 | 1.801 | 0.143 |
| Cyclic Experiment II run 2 | 1.847 | 0.106 |
| Cyclic Experiment II run 3 | 2.019 | 0.131 |

**Table S2.** Compiled data for the cyclic dissolution experiments. The numbers show $Ca^{2+}$ concentration from AAS analysis (mg/L).

| | | PM1 Cycles | | | | | | PM2 Cycles1 | | | | | | Chalk Cycle1 | |
|---|---|---|---|---|---|---|---|---|---|---|---|---|---|---|---|
| | TIME | Repl 1 | Repl1 | Repl 2 | Repl2 | Repl 3 | Repl3 | Repl 1 | Repl1 | Repl 2 | Repl2 | Repl 3 | Repl3 | | |
| Sample Name | | Dil 50 | Dil100 | Dil 50 | Dil100 | Dil 50 | Dil100 | Dil 50 | Dil100 | Dil 50 | Dil100 | Dil 50 | Dil100 | Average | STDEV |
| | 1 | | | | | | | | | | | | | | |
| cycle1sample1a | 2 | 236.5 | | 237.2 | | 238.9 | | 248.4991 | | 248.827 | | 249.1016 | | 243.2 | 5.67891 |
| | 3 | | | | | | | | | | | | | | |
| | 4 | | | | | | | | | | | | | | |
| cycle1sample1b | 5 | 301.0 | 321.3 | 300.9 | 320.3 | 299.6 | 319.4 | 316.6094 | | 316.4085 | | 313.5685 | | 312.1 | 8.502598 |
| cycle1sample1c | 10 | 320.3 | 336.3 | 321.7 | 335.1 | 321.1 | 336.1 | 342.3224 | | 341.8244 | | 341.8187 | | 332.9 | 8.800875 |
| cycle1sample1d | 20 | | 355.1 | | 355.3 | | 354.2 | | 379.6151 | | 376.9891 | | 377.642 | 366.5 | 11.64287 |
| cycle1sample1e | 35 | | 363.0 | | 364.3 | | 365.2 | | 388.0297 | | 388.265 | | 389.7182 | 376.4 | 12.28003 |
| cycle1sample1f | 65 | | 370.0 | | 372.6 | | 368.6 | | 392.4224 | | 392.7433 | | 394.0092 | 381.7 | 11.39279 |
| cycle1sample1g | 125 | | 375.5 | | 374.6 | | 375.9 | | 399.3743 | | 399.8947 | | 399.2393 | 387.4 | 12.08717 |
| cycle1sample1h | 215 | | 374.7 | | 374.5 | | 373.3 | | 395.3032 | | 396.236 | | 399.3522 | 385.6 | 11.46928 |
| cycle1sample1i | 335 | | 376.0 | | 374.6 | | 373.9 | | 397.3021 | | 397.6552 | | 397.3895 | 386.1 | 11.32183 |
| cycle1sample1j | 515 | | 371.8 | | 374.1 | | 376.8 | | 403.9465 | | 403.5861 | | 402.0212 | 388.7 | 14.55567 |
| cycle1sample1k | 1260 | | 372.0 | | 373.1 | | 372.8 | | 401.4322 | | 399.2167 | | 401.5255 | 386.7 | 14.08127 |
| cycle1sample1l | 1560 | | 378.1 | | 378.4 | | 376.5 | | 397.5161 | | 397.7548 | | 398.0276 | 387.7 | 10.06751 |
| cycle1sample1m | 1860 | | 373.2 | | 373.9 | | 373.5 | | 399.8015 | | 400.7559 | | 399.8895 | 386.8 | 13.31196 |
| cycle1sample1n | 2700 | | 372.1 | | 372.2 | | 370.1 | | 398.5674 | | 400.572 | | 400.979 | 385.8 | 14.32097 |
| cycle1sample1o1 | 3000 | | 370.5 | | 369.4 | | 371.2 | | 395.7119 | | 396.0862 | | 396.5704 | 383.3 | 12.87914 |
| cycle1sample1o2 | 3000 | | 370.6 | | 370.0 | | 368.0 | | 400.9827 | | 397.4639 | | 398.6903 | 384.3 | 14.81524 |
| cycle1sample1o3 | 3000 | | 371.5 | | 372.3 | | 370.2 | | 404.08 | | 405.0614 | | 401.8308 | 387.5 | 16.19783 |
| cycle1sample1p | 3300 | | 372.0 | | 374.0 | | 372.3 | | 397.4335 | | 397.398 | | 397.7355 | 385.1 | 12.3972 |

(Table S2 Continued)

| Sample Name | TIME | Repl 1 Dil 50 | Repl1 Dil100 | Repl 2 Dil 50 | Repl2 Dil100 | Repl 3 Dil 50 | Repl3 Dil100 | Repl 1 Dil 50 | Repl1 Dil100 | Repl 2 Dil 50 | Repl2 Dil100 | Repl 3 Dil 50 | Repl3 Dil100 | Chalk Cycle2 Average | STDEV |
|---|---|---|---|---|---|---|---|---|---|---|---|---|---|---|---|
| | 1 | | | | | | | | | | | | | | |
| cycle2sample1a | 2 | 223.9802 | | 225.1288 | | 223.3555 | | 235.5176 | | 235.9624 | | 235.3318 | | 229.9 | 5.751102 |
| | 3 | | | | | | | | | | | | | | |
| | 4 | | | | | | | | | | | | | | |
| cycle2sample1b | 5 | 283.3363 | 287.9682 | 284.1577 | 286.1884 | 286.1871 | 283.0675 | 287.1809 | | 286.6513 | | 287.5316 | | 285.8 | 1.727707 |
| cycle2sample1c | 10 | 306.3634 | 315.5103 | 311.4997 | 317.6849 | 312.5401 | 312.5126 | 306.8475 | 319.0839 | 308.3214 | 320.0626 | 310.0575 | 321.1517 | 313.5 | 4.950518 |
| cycle2sample1d | 20 | | 329.421 | | 333.9181 | | 333.8097 | | 331.4746 | | 333.9269 | | 332.644 | 332.5 | 1.64866 |
| cycle2sample1e | 35 | | 331.1485 | | 337.0874 | | 338.2212 | | 342.2796 | | 338.8056 | | 342.4457 | 338.3 | 3.783612 |
| cycle2sample1f | 65 | | 343.9181 | | 348.1967 | | 342.11 | | 346.9592 | | 348.1454 | | 348.4872 | 346.3 | 2.43074 |
| cycle2sample1g | 125 | | 351.4604 | | 348.8742 | | 348.2561 | | 355.4763 | | 354.9886 | | 350.9772 | 351.7 | 2.753664 |
| cycle2sample1h | 215 | | 354.1675 | | 352.3424 | | 352.4707 | | 352.9016 | | 353.5936 | | 355.1945 | 353.4 | 1.005417 |
| cycle2sample1i | 335 | | 351.5429 | | 356.2609 | | 356.5602 | | 357.2711 | | 358.0591 | | 357.1559 | 356.1 | 2.133824 |
| cycle2sample1j | 515 | | 360.9467 | | 362.0782 | | 358.5573 | | 351.0752 | | 352.3179 | | 352.5946 | 356.3 | 4.414973 |
| cycle2sample1k | 1260 | | 358.0294 | | 357.4566 | | 357.814 | | 358.1515 | | 358.0675 | | 355.5378 | 357.5 | 0.91078 |
| cycle2sample1l | 1560 | | 357.4715 | | 355.5343 | | 352.4445 | | 358.1277 | | 359.9663 | | 360.4903 | 357.3 | 2.729022 |
| cycle2sample1m | 1860 | | 355.0903 | | 355.9439 | | 350.716 | | 358.1678 | | 359.7745 | | 361.3032 | 356.8 | 3.45597 |
| cycle2sample1n | 2700 | | 359.2323 | | 353.688 | | 355.2423 | | 362.0592 | | 361.6784 | | 361.3235 | 358.9 | 3.271668 |
| cycle2sample1o1 | 3000 | | 349.9641 | | 357.5124 | | 355.4246 | | 359.4416 | | 358.9935 | | 360.7459 | 357.0 | 3.564042 |
| cycle2sample1o2 | 3000 | | 351.6135 | | 361.0654 | | 360.4456 | | 362.7392 | | 362.0542 | | 363.7245 | 360.3 | 4.016888 |
| cycle2sample1o3 | 3000 | | 362.5665 | | 367.1028 | | 365.2859 | | 364.6778 | | 363.9955 | | 361.0596 | 364.1 | 1.930567 |
| cycle2sample1p | 3300 | | 360.4872 | | 361.4375 | | 363.5464 | | 358.3314 | | 355.5844 | | 354.443 | 359.0 | 3.20671 |

(Table S2 Continued)

| Sample Name | TIME | Repl 1 Dil 50 | Repl1 Dil100 | Repl 2 Dil 50 | Repl2 Dil100 | Repl 3 Dil 50 | Repl3 Dil100 | Repl 1 Dil 50 | Repl1 Dil100 | Repl 2 Dil 50 | Repl2 Dil100 | Repl 3 Dil 50 | Repl3 Dil100 | Chalk Cycle3 Average | STDEV |
|---|---|---|---|---|---|---|---|---|---|---|---|---|---|---|---|
| cycle3sample1a | 2 | 155.4149 | | 155.2155 | | 156.5692 | | 169.4929 | | 169.6111 | | 169.9187 | | 162.7 | 6.984434 |
| cycle3sample1b | 5 | 211.1219 | 227.9583 | 211.2215 | 227.6317 | 211.6111 | 227.5764 | 229.9356 | | 229.6919 | | 228.4813 | | 222.8 | 8.158776 |
| cycle3sample1c | 10 | 258.7844 | 271.7913 | 257.7132 | 272.0481 | 258.6115 | 270.6978 | 268.5991 | 274.0544 | 269.8973 | 272.2265 | 268.138 | 275.7923 | 268.2 | 6.025515 |
| cycle3sample1d | 20 | | 305.201 | | 302.7105 | | 303.0639 | | 301.8844 | | 301.7306 | | 303.1737 | 303.0 | 1.140841 |
| cycle3sample1e | 35 | | 317.0673 | | 316.1668 | | 313.901 | | 311.761 | | 312.8086 | | 312.6989 | 314.1 | 1.924042 |
| cycle3sample1f | 65 | | 321.707 | | 322.7661 | | 322.3918 | | 317.9488 | | 323.4982 | | 320.6976 | 321.5 | 1.811488 |
| cycle3sample1g | 125 | | 322.3265 | | 323.4667 | | 324.7246 | | 323.0256 | | 324.5205 | | 323.6422 | 323.6 | 0.824778 |
| cycle3sample1h | 215 | | 323.6408 | | 324.1841 | | 324.7552 | | 331.5623 | | 327.9622 | | 332.6738 | 327.5 | 3.579996 |
| cycle3sample1i | 335 | | 323.6858 | | 329.3566 | | 327.0239 | | 330.0668 | | 328.6881 | | 328.2285 | 327.8 | 2.083751 |
| cycle3sample1j | 515 | | 329.4013 | | 326.6536 | | 327.9746 | | 330.4269 | | 333.3852 | | 331.7313 | 329.9 | 2.24603 |
| cycle3sample1k | 1260 | | 333.8516 | | 337.5898 | | 337.6086 | | 339.7906 | | 339.229 | | 338.5632 | 337.8 | 1.925851 |
| cycle3sample1l | 1560 | | 330.4339 | | 333.0889 | | 329.7484 | | 337.4805 | | 337.1057 | | 336.3783 | 334.0 | 3.136585 |
| cycle3sample1m | 1860 | | 335.1483 | | 338.697 | | 336.9338 | | 337.8786 | | 337.8248 | | 338.1813 | 337.4 | 1.153124 |
| cycle3sample1n | 2700 | | 336.5192 | | 336.5206 | | 337.7925 | | 337.7638 | | 340.2544 | | 339.4337 | 338.0 | 1.390738 |
| cycle3sample1o1 | 3000 | | 338.5111 | | 336.9101 | | 336.0752 | | 342.4737 | | 342.8484 | | 342.1938 | 339.8 | 2.770405 |
| cycle3sample1o2 | 3000 | | 339.0031 | | 339.1042 | | 339.6755 | | 339.3754 | | 341.4798 | | 342.8681 | 340.3 | 1.433481 |
| cycle3sample1o3 | 3000 | | 340.2957 | | 339.0893 | | 341.1799 | | 339.0322 | | 335.6326 | | 338.0981 | 338.9 | 1.756754 |
| cycle3sample1p | 3300 | | 338.1778 | | 336.8439 | | 340.1223 | | 346.6544 | | 344.6276 | | 343.1446 | 341.6 | 3.502742 |

(Table S2 Continued)

| Sample Name | TIME | cycle1 LabCal1 | | | | | | LabCal2 cycle 1 | | | | | | LabCal Cycle1 | |
|---|---|---|---|---|---|---|---|---|---|---|---|---|---|---|---|
| | | Repl 1 | Repl1 | Repl 2 | Repl2 | Repl 3 | Repl3 | Repl 1 | Repl1 | Repl 2 | Repl2 | Repl 3 | Repl3 | | |
| | | Dil 50 | Dil100 | Dil 50 | Dil100 | Dil 50 | Dil100 | Dil 50 | Dil100 | Dil 50 | Dil100 | Dil 50 | Dil100 | Average | STDEV |
| | 1 | | | | | | | | | | | | | | |
| cycle1sample1a | 2 | 117.0 | | 117.4 | | 117.5 | | 114.1881 | | 114.1412 | | 114.4156 | | 115.8 | 1.54078 |
| | 3 | | | | | | | | | | | | | | |
| | 4 | | | | | | | | | | | | | | |
| cycle1sample1b | 5 | 202.1 | 215.6 | 202.3 | 216.7 | 201.9 | 216.6 | 203.305 | | 202.3433 | | 201.3721 | | 206.9 | 6.668224 |
| cycle1sample1c | 10 | 268.2 | 279.4 | 268.0 | 283.0 | 269.3 | 282.3 | 268.2222 | 284.8826 | 267.0254 | 285.5384 | 268.2601 | 285.2026 | 275.8 | 7.762951 |
| cycle1sample1d | 20 | | 318.7 | | 320.7 | | 320.6 | | 344.0722 | | 345.4118 | | 344.2296 | 332.3 | 12.30929 |
| cycle1sample1e | 35 | | 333.3 | | 332.9 | | 335.2 | | 354.9202 | | 352.9314 | | 355.1351 | 344.1 | 10.32288 |
| cycle1sample1f | 65 | | 340.4 | | 336.8 | | 340.4 | | 362.5043 | | 365.032 | | 362.7792 | 351.3 | 12.21221 |
| cycle1sample1g | 125 | | 342.6 | | 340.4 | | 341.8 | | 362.7157 | | 363.6664 | | 364.4925 | 352.6 | 11.04102 |
| cycle1sample1h | 215 | | 332.7 | | 331.7 | | 333.1 | | 364.7518 | | 364.5975 | | 364.7576 | 348.6 | 16.09929 |
| cycle1sample1i | 335 | | 332.2 | | 335.2 | | 335.5 | | 364.4831 | | 363.6125 | | 365.3202 | 349.4 | 15.13448 |
| cycle1sample1j | 515 | | 335.9 | | 336.8 | | 338.2 | | 363.6461 | | 366.4516 | | 363.6387 | 350.8 | 13.84612 |
| cycle1sample1k | 1260 | | 342.6 | | 341.5 | | 344.1 | | 359.6521 | | 360.6965 | | 360.557 | 351.5 | 8.81898 |
| cycle1sample1l | 1560 | | 342.4 | | 342.4 | | 343.7 | | 364.2836 | | 364.4668 | | 365.3992 | 353.8 | 10.94741 |
| cycle1sample1m | 1860 | | 341.5 | | 340.3 | | 341.8 | | 365.9859 | | 365.0785 | | 365.4004 | 353.3 | 12.15298 |
| cycle1sample1n | 2700 | | 337.3 | | 336.8 | | 339.4 | | 364.6251 | | 362.1724 | | 364.2848 | 350.8 | 12.9708 |
| cycle1sample1o1 | 3000 | | 338.8 | | 339.3 | | 339.1 | | 361.594 | | 362.6263 | | 360.1428 | 350.3 | 11.22411 |
| cycle1sample1o2 | 3000 | | 339.4 | | 338.7 | | 340.0 | | 362.9593 | | 365.7282 | | 366.237 | 352.2 | 12.84646 |
| cycle1sample1o3 | 3000 | | 338.8 | | 340.0 | | 339.1 | | 363.9811 | | 364.9126 | | 364.7663 | 351.9 | 12.64138 |
| cycle1sample1p | 3300 | | 337.6 | | 339.5 | | 341.8 | | 354.8037 | | 356.4835 | | 356.8268 | 347.8 | 8.313916 |

(Table S2 Continued)

| Sample Name | TIME | cycle2 LabCal1 | | | | | | cycle2 LabCal2 | | | | | | LabCal Cycle2 | |
| --- | --- | --- | --- | --- | --- | --- | --- | --- | --- | --- | --- | --- | --- | --- | --- |
| | | Repl 1 | Repl1 | Repl 2 | Repl2 | Repl 3 | Repl3 | Repl 1 | Repl1 | Repl 2 | Repl2 | Repl 3 | Repl3 | | |
| | | Dil 50 | Dil100 | Dil 50 | Dil100 | Dil 50 | Dil100 | Dil 50 | Dil100 | Dil 50 | Dil100 | Dil 50 | Dil100 | Average | STDEV |
| | 1 | | | | | | | | | | | | | | |
| cycle2sample1a | 2 | 84.52583 | | 87.09012 | | 87.96206 | | | | | | | | 86.5 | 1.458447 |
| | 3 | | | | | | | | | | | | | | |
| | 4 | | | | | | | | | | | | | | |
| cycle2sample1b | 5 | 170.3322 | 158.2221 | 171.6143 | 160.8595 | 171.7122 | 157.1492 | 126.6215 | | 127.1318 | | 127.2379 | | 152.3 | 18.6423 |
| cycle2sample1c | 10 | 249.3995 | 246.2513 | 250.2134 | 246.0506 | 246.5014 | 242.7405 | 194.4784 | 203.7514 | 194.4114 | 204.5338 | 193.3555 | 203.8096 | 223.0 | 24.22259 |
| cycle2sample1d | 20 | | 313.0348 | | 309.4319 | | 309.4334 | | 278.839 | | 279.193 | | 279.554 | 294.9 | 15.76616 |
| cycle2sample1e | 35 | | 339.6891 | | 339.8475 | | 336.7449 | | 325.7322 | | 328.0197 | | 326.1613 | 332.7 | 6.184748 |
| cycle2sample1f | 65 | | 354.5582 | | 355.0274 | | 350.6467 | | 353.6772 | | 353.4934 | | 353.1825 | 353.4 | 1.396289 |
| cycle2sample1g | 125 | | 358.4846 | | 358.1194 | | 357.3817 | | 360.6462 | | 361.6486 | | 362.7682 | 359.8 | 1.972128 |
| cycle2sample1h | 215 | | 358.1941 | | 351.1224 | | 353.2983 | | 364.8734 | | 367.7946 | | 365.4936 | 360.1 | 6.345223 |
| cycle2sample1i | 335 | | 355.0213 | | 356.0171 | | 357.8373 | | 368.6213 | | 367.3586 | | 369.887 | 362.5 | 6.26275 |
| cycle2sample1j | 515 | | 354.2898 | | 357.384 | | 357.7482 | | 363.6305 | | 362.1592 | | 364.0812 | 359.9 | 3.627116 |
| cycle2sample1k | 1260 | | 359.4751 | | 360.5875 | | 361.5094 | | 361.6766 | | 361.6816 | | 361.5349 | 361.1 | 0.809048 |
| cycle2sample1l | 1560 | | 355.8984 | | 357.5745 | | 356.7224 | | 362.7421 | | 361.0737 | | 359.6022 | 358.9 | 2.431758 |
| cycle2sample1m | 1860 | | 359.0952 | | 356.3708 | | 359.0785 | | 361.0869 | | 361.1917 | | 361.4843 | 359.7 | 1.787283 |
| cycle2sample1n | 2700 | | 362.9952 | | 359.4718 | | 361.9946 | | 361.8668 | | 361.0259 | | 361.1354 | 361.4 | 1.083427 |
| cycle2sample1o1 | 3000 | | 354.1745 | | 353.1046 | | 354.3484 | | 368.8886 | | 365.8291 | | 362.2527 | 359.8 | 6.206927 |
| cycle2sample1o2 | 3000 | | 351.3057 | | 358.3693 | | 352.5242 | | 367.2789 | | 370.7772 | | 370.594 | 361.8 | 8.122858 |
| cycle2sample1o3 | 3000 | | 357.573 | | 352.5578 | | 353.7782 | | 368.4427 | | 365.8926 | | 367.4911 | 361.0 | 6.539919 |
| cycle2sample1p | 3300 | | 357.4133 | | 356.7101 | | 358.6231 | | 365.1308 | | 367.6898 | | 365.6951 | 361.9 | 4.400047 |

(Table S2 Continued)

| Sample Name | TIME | cycle3 LabCal1 | | | | | | cycle3 LabCal2 | | | | | | LabCal Cycle3 | |
|---|---|---|---|---|---|---|---|---|---|---|---|---|---|---|---|
| | | Repl 1 | Repl1 | Repl 2 | Repl2 | Repl 3 | Repl3 | Repl 1 | Repl1 | Repl 2 | Repl2 | Repl 3 | Repl3 | | |
| | | Dil 50 | Dil100 | Dil 50 | Dil100 | Dil 50 | Dil100 | Dil 50 | Dil100 | Dil 50 | Dil100 | Dil 50 | Dil100 | Average | STDEV |
| | 1 | | | | | | | | | | | | | | |
| cycle3sample1a | 2 | -1.79499 | | -1.97187 | | -1.09063 | | 42.03781 | | 42.1814 | | 41.53073 | | 20.1 | 21.77046 |
| | 3 | | | | | | | | | | | | | | |
| | 4 | | | | | | | | | | | | | | |
| cycle3sample1b | 5 | 5.828418 | | 5.482213 | | 6.069461 | | 79.12391 | | 79.34334 | | 80.13236 | | 42.7 | 36.87159 |
| cycle3sample1c | 10 | 188.3228 | | 190.2074 | | 191.2146 | | 124.6898 | 123.3487 | 124.8147 | 124.5319 | 125.2177 | 123.5504 | 146.2 | 30.91613 |
| cycle3sample1d | 20 | | 152.8846 | | 150.7856 | | 148.364 | | 188.2222 | | 188.6665 | | 188.6043 | 169.6 | 18.95538 |
| cycle3sample1e | 35 | | 224.3658 | | 222.3287 | | 223.4531 | | 249.3194 | | 248.0746 | | 250.2671 | 236.3 | 12.9479 |
| cycle3sample1f | 65 | | 289.3735 | | 294.4267 | | 290.3147 | | 294.716 | | 293.39 | | 296.3958 | 293.1 | 2.482014 |
| cycle3sample1g | 125 | | 306.8385 | | 306.1526 | | 307.4859 | | 327.7339 | | 327.7783 | | 328.7993 | 317.5 | 10.65172 |
| cycle3sample1h | 215 | | 319.4784 | | 321.4999 | | 318.3109 | | 332.7516 | | 334.3642 | | 336.8911 | 327.2 | 7.606921 |
| cycle3sample1i | 335 | | 322.4867 | | 322.3862 | | 324.0572 | | 328.283 | | 328.731 | | 328.26 | 325.7 | 2.781442 |
| cycle3sample1j | 515 | | 315.9189 | | 317.0482 | | 314.5126 | | 335.0605 | | 331.6395 | | 336.0822 | 325.0 | 9.343279 |
| cycle3sample1k | 1260 | | 335.5559 | | 337.6188 | | 337.7191 | | 327.4518 | | 326.9896 | | 329.8896 | 332.5 | 4.572272 |
| cycle3sample1l | 1560 | | 324.4207 | | 325.0041 | | 325.4394 | | 340.8546 | | 338.023 | | 341.2587 | 332.5 | 7.619403 |
| cycle3sample1m | 1860 | | 323.6295 | | 326.2601 | | 328.4763 | | 345.2198 | | 343.9956 | | 345.5048 | 335.5 | 9.507532 |
| cycle3sample1n | 2700 | | 324.2375 | | 322.9594 | | 323.2965 | | 340.5164 | | 342.2337 | | 343.5649 | 332.8 | 9.353168 |
| cycle3sample1o1 | 3000 | | 316.0004 | | 317.4729 | | 316.5977 | | 336.9335 | | 343.3385 | | 342.809 | 328.9 | 12.34761 |
| cycle3sample1o2 | 3000 | | 319.0747 | | 319.8429 | | 321.7304 | | 344.2212 | | 347.7713 | | 346.0512 | 333.1 | 12.96397 |
| cycle3sample1o3 | 3000 | | 322.2946 | | 324.3924 | | 321.5886 | | 344.458 | | 346.8083 | | 344.383 | 334.0 | 11.28859 |
| cycle3sample1p | 3300 | | 319.6139 | | 322.0358 | | 320.2777 | | 322.0406 | | 323.4135 | | 321.565 | 321.5 | 1.244727 |